\documentclass[prb,aps,twocolumn,showpacs]{revtex4}

\usepackage{graphicx}
\usepackage{dcolumn}
\usepackage{bm}

\begin{document}

\title{Collective Spin Dynamics in the "Coherence Window" for Quantum Nanomagnets}

\author{I. S. Tupitsyn$^{1,2}$}
\affiliation{$^1$ Pacific Institute for Theoretical Physics, University of
British Columbia, \\
6224 Agricultural Rd., Vancouver, B.C. V6T 1Z1, Canada \\
$^2$ Theoretical Division of the Institute of Superconductivity and
Solid State Physics, \\
Russian Research Centre "Kurchatov Institute", Kurchatov Sq.1, Moscow 123182,
Russia}

\begin{abstract}

The spin coherence phenomena and the possibility of their
observation in nanomagnetic insulators attract more and more
attention in the last several years. Recently it has been shown that
in these systems in large transverse magnetic field there can be a
fairly narrow "coherence window" for phonon and nuclear
spin-mediated decoherence. What kind of spin dynamics can then be
expected in this window in a crystal of magnetic nanomolecules
coupled to phonons, to nuclear spin bath and {\it to each other via
dipole-dipole interactions}? Studying multispin correlations,
we determine the region of parameters where "coherent clusters" of
collective spin excitations can appear. Although two particular
systems, namely crystals of $Fe_8$-triazacyclonane and
$Mn_{12}$-acetate molecules, are used in this work to illustrate the
results, here we are not trying to predict an existence of
collective coherent dynamics in some particular system. Instead, we discuss
the way how any crystalline system of dipole-dipole coupled nanomolecules
can be analyzed to decide whether this system is suitable for attempts to
observe coherent dynamics. The presented analysis can be useful
in the search for magnetic systems showing the spin coherence phenomena.

\end{abstract}

\maketitle

\vspace{1cm}


\section{Introduction}
\label{sec:intr}

In the last decade the quantum tunneling phenomenon in nanomagnetic
insulators has been attracting extensive interest. Many experiments have
been done to study the tunneling relaxation in the ensembles of magnetic
molecules with central molecular spins $\vec{S}_i$. \cite{can00,BB00,WWrev,TB}
These molecules couple to each other via dipole-dipole interactions,
\cite{PSPRL,VillSurf,FernRelax,TSDIP} to phonons
\cite{POL96,LuisBart,LossLeu,Pohjola,TB} and to nuclear spins.
\cite{PSPRL,PS96,PSSB} The early study of these systems in a low temperature
regime has been concentrated mainly on the incoherent tunneling in low
transverse fields, when the magnitude of the ground state tunneling
splitting $2\Delta_o$ (produced by the tunneling between two potential
wells separated by a barrier of magnetic anisotropy; $\Delta_o$ is the
tunneling matrix element) is small in comparison with the parameters
describing interactions with environment providing anomalously high
decoherence. During last several years more attention has been paid to the
spin coherence phenomena. \cite{SpinPrec,AepSci,LossCoh,Kent,Chud04}

As it has been shown recently, \cite{STPRB} in nanomagnetic insulators in
large transverse fields, where $\Delta_o(H^{\perp})$ increases, there
can be a field region ("coherence window") in which both phonon and
nuclear spin-mediated decoherence are drastically reduced (electronic
decoherence in magnetic insulators is absent). The existence of such
coherence window is important both for fundamental physics (attempts
to find materials showing coherent spin tunneling phenomenon) and for
quantum device engineering (attempts to make a solid-state qubit).

At very low temperature each molecule with large central spin
$\vec{S}_i$ can be modelled as a two-level system (Appendix \ref{sub:apA}),
whose Hamiltonian $H_i = -\Delta_i \hat{\tau}^x_i - \xi_i \hat{\tau}^z_i$
operates in a subspace of only two lowest states of $\vec{S}_i$. With
$\Delta_i$ and $\xi_i$ being the ground state tunneling matrix element
and the longitudinal bias acting on $i$-th molecule, this two-state
representation is valid only if $\Delta_i$ is small in comparison with the
spin gap $E_g$ to the next levels. For example, in two well-known central spin
$|\vec{S}|=10$ systems, $Fe_8$-triazacyclonane ($Fe_8$) and $Mn_{12}$-acetate
($Mn_{12}$), this condition is met since $E_g \sim 5 \; K$ and $\sim 11 \; K$
respectively while the values of a zero-field tunneling splitting are $\sim
10^{-7} \; K$ and $\sim 10^{-11} \; K$.

Suppose that molecules do not interact with each other. Then
the central spin of any molecule can oscillate between states
$|\uparrow \rangle$ and $|\downarrow \rangle$ (\ref{TLS.3})
and this process is described by the probability
$P_{\downarrow \uparrow}(t)$ (\ref{TLS.4}). If $\Delta >> \xi$,
the amplitude of these oscillations is $\approx 1$. If the central
spin of each molecule is also isolated from its nuclear subsystem
and from the phonon thermostat, the tunneling oscillations, being
coherent, can last for an infinitely long time. Interactions with
the nuclear spin and the phonon thermostats lead to decoherence and,
after the so-called decoherence time $\tau_{\phi}$, coherence will
be suppressed and oscillations will disappear.

The decoherence "quality factor", giving an estimation for the number of
coherent oscillations in the system before coherence will be suppressed, is
$Q_{\phi} \sim 1 / \gamma_{\phi}$, where $\gamma_{\phi} = \hbar / (\Delta_o
\tau_{\phi})$ is the dimensionless decoherence rate. The contributions to the
decoherence time $\tau_{\phi}$ from interactions with the nuclear spins and
phonons ($\tau^{nu}_{\phi}$ and $\tau^{ph}_{\phi}$, respectively) are:
\cite{STPRB,STChPh}
\begin{equation}
{ 1 \over \tau^{nu}_{\phi}} = { E^2_o \over 2 \Delta_o \hbar  }; \;\;\;
{ 1 \over \tau^{ph}_{\phi}} = {S^2 \Omega^2_o \Delta_o^3 \over \Theta_D^4
\hbar} \coth(\Delta_o / k_B T),
\label{PH_NU}
\end{equation}
where $E_o$ is the half-width of the Gaussian distribution of the hyperfine
bias energies; $\Theta_D$ is the Debye energy; and $\Omega_o \sim E_g$ is the
energy of small oscillations in the potential wells.

The goal of the present work is to study the spin dynamics in ensembles
of {\it dipole-dipole coupled magnetic molecules} in the coherence
window for the nuclear spin and phonon degrees of freedom at times
$t < \tau_{np}$,
\begin{equation}
\tau_{np}= \min \{ \tau^{nu}_{\phi}, \tau^{ph}_{\phi} \}.
\label{taunp}
\end{equation}
Namely, in this work we would like to study the internal dynamics of a
{\it temperature equilibrated system}, but not the dynamics {\it induced}
by the artificial preparation of a system at $t=0$, say, in state
$|\uparrow \uparrow \ldots \uparrow \rangle$ (the analysis of the latter
problem will be presented separately). From now on, for the sake of brevity,
the coherence window will be called the {\it NPC-window} (nuclear spins and
phonons coherence window). To illustrate the results, all particular
calculations will be based on the parameters for two systems, namely,
for crystals of $Fe_8$ and $Mn_{12}$ molecules.

\section{Hamiltonian and interactions}
\label{sec:model}

At very low temperatures a set of molecules with central molecular
spins $|\vec{S}_i| = S$ coupled to each other via the dipole-dipole
interaction can be described by the effective Hamiltonian:
\begin{equation}
H = \sum_i (-\Delta_i \hat{\tau}^x_i - \xi^{en}_i \hat{\tau}^z_i)
+ {1 \over 2} \sum_{i j} \hat{V}_{dd}(\vec{r}_{ij}),
\label{EffH}
\end{equation}
where $\hat{\tau}^z$ and $\hat{\tau}^x$ are the Pauli matrixes; $\Delta_i$
is the tunneling matrix element; and $\xi^{en}_i$ is the bias acting on
$i$-th molecule from external and nuclear fields. The last term in
(\ref{EffH}) describes the dipolar coupling between pairs of molecules,
separated by distance $|\vec{r}_{ij}|=|\vec{r}_i - \vec{r}_j|$:
\begin{equation}
\hat{V}_{dd}(\vec{r}_{ij}) = {E_D \over |\vec{r}_{ij}|^3} \left(
\hat{\vec{\tau}}_i \hat{\vec{\tau}}_j -
3 {(\hat{\vec{\tau}}_i \vec{r}_{ij}) (\hat{\vec{\tau}}_j \vec{r}_{ij})
\over |\vec{r}_{ij}|^2 } \right),
\label{DI}
\end{equation}
where $E_D = (\mu_0 / 4 \pi) g^2_e \mu^2_B S^2$; $\mu_0 / 4\pi = 10^{-7} N/A^2$
(in the SI system of units); $g_e$ is the electronic $g$-factor; and $\mu_B$ is
the Bohr magneton. Note that Hamiltonian (\ref{EffH}) does not include
the interactions with phonons and nuclear spins. Instead, the known results
\cite{STPRB,STChPh} for the phonon and nuclear spin decoherence rates,
Eq.(\ref{PH_NU}), will be used.

Coherence window for the nuclear spin and phonon channels of decoherence
opens up at high transverse fields, where the value of the tunneling splitting
becomes large in comparison with the parameters describing interactions of
the central spin with the environment. At these conditions all $\Delta_i$ in
a sample are approximately the same \cite{DeltaDistr} and for brevity
can be replaced (where it is reasonable) by one parameter $\Delta_o$, whose
transverse field dependence $\Delta_o(\vec{H}^{\perp})$ can be calculated
using the corresponding molecular Hamiltonian for the central spin $\vec{S}$.
For both the $Fe_8$ and the $Mn_{12}$ molecules these Hamiltonians are
(approximately) known.

\emph{\textbf{(1) The "central spin" Hamiltonians for $Fe_8$ and $Mn_{12}$
molecules.}} ~ Below $\sim 10 \; K$ for $Fe_8$ and below $\sim 40 \; K$ for
$Mn_{12}$ these molecules are described by two similar $S=10$ Hamiltonians
of magnetic anisotropy:
\begin{equation}
H^{(Fe)}_S = -D S^2_z + E S^2_x + K^{\perp}_4 (S^4_{+}+S^4_{-})
- g_e \mu_B \vec{H} \vec{S},
\label{Hfe}
\end{equation}
with \cite{FeExpl} $D/k_B = 0.23 \; K$, $E/k_B = 0.094 \; K$,
and $K_4/k_B = -3.28 \times 10^{-5} \; K$; and
\begin{equation}
H^{(Mn)}_S=-D S^2_z - K^{||}_4 S^4_z + K^{\perp}_4 (S^4_{+}+S^4_{-})
- g_e \mu_B \vec{H} \vec{S},
\label{Hmn}
\end{equation}
with \cite{Mireb} $D/k_B = 0.548 \; K$, $K^{||}_4/k_B = 1.17 \times 10^{-3}
\; K$, $K^{\perp}_4/k_B=2.2 \times 10^{-5} \; K$.

Note that in the $Fe_8$ system the tunneling splitting
$\Delta_o(\vec{H}^{\perp})$ and its period of oscillations with
$\vec{H}^{\perp}$ have been measured \cite{WWPRL,WWSSCI} while in the
$Mn_{12}$ system these parameters have never been measured. The latter
makes it rather problematic to verify the value of the tunneling splitting
obtained directly from the Hamiltonian (\ref{Hmn}). However, we would
like to study the region of large transverse fields where
$\Delta_o(\vec{H}^{\perp})$ is already large (although $\Delta_o << E_g$)
and is less sensitive to some variations of the anisotropy constants
\cite{Mn2ndOrd} (moreover, at some stage we start to make estimations
rather than exact calculations). Thus, in what follows we use the
Hamiltonians (\ref{Hfe}) and (\ref{Hmn}) for $Fe_8$ and $Mn_{12}$ molecules.

\emph{\textbf{(2) Dipolar interactions.}} ~ For the sake of definiteness we
apply a transverse magnetic field along the $x$-axis, so that only the
$S^z_i$ and the $S^x_i$ projections of the total molecular spin $\vec{S}_i$
are nonzero. Therefore, the interaction term $\hat{V}_{dd}$ can be rewritten
as:
\begin{equation}
\hat{V}_{dd} (\vec{r}_{ij}) = \sum_{\{ \alpha, \beta \} = \{x,z\}}
V^{\alpha \beta}_{dd} (\vec{r}_{ij}) \hat{\tau}^{\alpha}_i \hat{\tau}^{\beta}_j,
\label{VDD.1}
\end{equation}
where all $V^{\alpha \beta}_{dd} (\vec{r}_{ij})$ can be obtained
from Eq.(\ref{DI}). The $i$-th bias energy $\xi^{en}_i$ in (\ref{EffH}),
as it is written, contains contributions only from the longitudinal
external and nuclear fields. The dipolar contribution to the total bias
$\xi_i$ acting on $i$-th molecule can be written in the form
$\xi^d_i=-g_e \mu_B S^z_i H^z_i(dip)$ and the longitudinal
dipolar filed $H^z_i(dip)$ at $i$-th site is:
\begin{equation}
\hat{H}^z_i(dip) = \sum_{j \ne i} {F_D \over |\vec{r}_{ij}|^3} \left(
3 {(\hat{\tau}^z_j z_{ij} + \hat{\tau}^{x}_j x_{ij}) z_{ij}
\over |\vec{r}_{ij}|^2 } - \hat{\tau}^z_j \right),
\label{VDD.2}
\end{equation}
where $F_D = (\mu_0 / 4 \pi) g_e \mu_B S$ and $z_{ij}, \; x_{ij}$ are the
corresponding components of vector $\vec{r}_{ij}$.

The distributions of the dipolar bias energies created by molecular
spins in polarized and depolarized samples are different in the
low transverse field limit and similar in the high transverse fields
limit (where $\vec{S}_i$ is oriented nearly along the transverse
field direction). At low transverse fields the half-width $W_D$ of
the dipolar bias distribution in a completely depolarized sample is
several times larger than in a polarized sample. At high transverse
fields the half-width $W_D$ in both samples is nearly the same
(comparing the longitudinal field distributions for polarized and
depolarized samples at $H^x = 4.8 \; T$ in Fig.\ref{fig:DE_Fig5} of
Appendix \ref{sub:Calc_meth} one can see that they are nearly the
same). This parameter can be calculated numerically for any sample.

\emph{\textbf{(3) Hyperfine interactions.}} ~ The interactions of
the central molecular spin $\vec{S}_i$ with the nuclear spin bath
lead to the "spread" of each molecular spin state characterized by
the half-width $E_o$ of the Gaussian distribution of the hyperfine
bias energies $\xi_N$. For $N_n$ nuclear spins $\vec{I}_k$ in
each molecule, one finds \cite{STPRB,STChPh,ROSE}
$E^2_o = \sum^{N_{n}}_{k=1} (I_k+1) I_k (\omega^{||}_k)^2 / 3$,
where $\{ \omega^{||}_k \}$ are the (longitudinal) couplings between
the central spin and each $k$-th nuclear spin. \cite{PS96,PSSB}
Knowledge of all nuclear moments and positions of all nuclei in the
molecule \cite{CDC} allows one to calculate all these coupling
constants and $E_o$. \cite{ROSE,STPRB,STChPh,TSDIP,WWISO}

\emph{\textbf{ (4) The transverse magnetic field behavior of
important parameters.}} ~ The ground state $| \Uparrow \rangle$
(symmetric) and the excited state $| \Downarrow \rangle$
(antisymmetric) (\ref{TLS.2}) of Hamiltonian (\ref{TLS.1}) are
separated by the energy gap $2 \varepsilon_i = 2 (\Delta^2_i +
\xi^2_i)^{1/2}$. At low temperatures ($k_B T < \Delta_o$) in
the limit $\Delta_i >> \xi_i$ in a temperature equilibrated sample
most of molecules are in states $|\Uparrow \rangle$. Then,
calculating matrix elements $\langle \Uparrow |\hat{\tau}^z_i|\Uparrow
\rangle = \xi_i / \varepsilon_i$ and $\langle \Uparrow |\hat{\tau}^x_i|
\Uparrow \rangle = \Delta_i / \varepsilon_i$, for $\Delta_i >> \xi_i$
one finds $\langle \Uparrow |S^z_i| \Uparrow \rangle \to 0$ and $\langle
\Uparrow | S^x_i | \Uparrow \rangle \to S$. Thus, as $\Delta_o$ increases
with the transverse field, both $W_D$ and $E_o$ should decrease.

\begin{figure}[ht]
\centering
\vspace{-1.8cm}
\hspace{-0.3cm}
\includegraphics[scale=0.38]{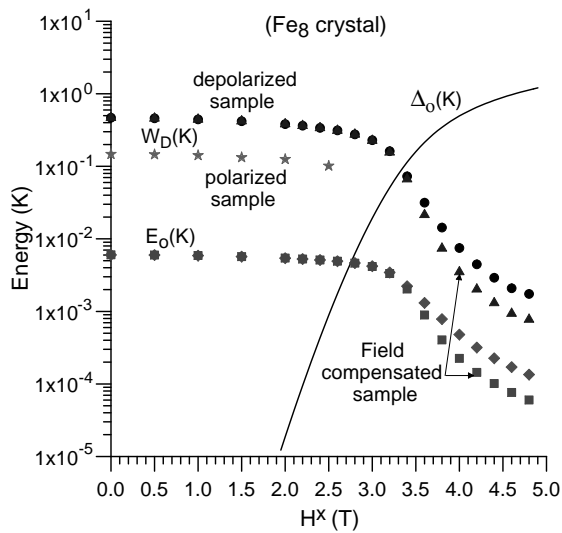}
\vspace{-1.7cm}
\caption{$Fe_8$ crystal (cluster of $50^3$ $Fe_8$ molecules, for
details see Appendix \ref{sub:Calc_meth}). The curves are: {\it circles}
- half-width $W_D$ of the dipolar bias distribution vs transverse
field $H^x$ (in Tesla) for model depolarized sample (see the text);
{\it stars} - $W_D(H^x)$ for polarized sample; {\it triangles} - same
as circles, but for the field-compensated sample; {\it diamonds}
- half-width $E_o$ of the hyperfine bias distribution; {\it squares}
- same as diamonds but for the field-compensated sample; {\it solid
line} - tunneling matrix element $\Delta_o(H^x)$. All molecules are in
a state $|\Uparrow \rangle$. ($W_D, \; E_o, \; \Delta_o$ are in Kelvins.)}
\label{fig:DE_Fig1}
\end{figure}

\vspace{-3mm}

When $\Delta_o << W_D$, in a depolarized sample the value of $W_D$ is
several times larger than in a polarized sample. To understand how
$W_D$ behaves at large transverse fields, it is sufficient to calculate
this parameter in a {\it model} depolarized sample where all molecules
are in states $|\Uparrow \rangle$, but $\sum_i S^z_i / |S^z_i| = 0$.
\cite{SxSign} The transverse field dependence of important parameters
for crystals of $Fe_8$ and $Mn_{12}$ molecules is presented in
Figs.\ref{fig:DE_Fig1} and \ref{fig:DE_Fig2} (the description of our
calculation procedure is given in Appendix \ref{sub:Calc_meth}).
\cite{RES_EX} Deviations from the results of Figs.\ref{fig:DE_Fig1}
and \ref{fig:DE_Fig2} for nonzero populations of states $|\Downarrow
\rangle$ are insignificant up to the limit of equipopulation - this is
clear, for example, from Fig.\ref{fig:DE_Fig5} (Appendix
\ref{sub:Calc_meth}).

Depending on the crystal structure and the sample geometry, the
dipolar fields distribution can be shifted (such a shift can be
rather large, see, for example, Fig.\ref{fig:DE_Fig5} in Appendix
\ref{sub:Calc_meth}). This shift changes with the transverse field.
The larger the shift, the slower both the $W_D(H^{\perp})$ and the
$E_o(H^{\perp})$ decrease with $H^{\perp}$. This effect can
be seen in the high-field part of Fig.\ref{fig:DE_Fig1}.
Two upper curves for both $W_D$ and $E_o$ represent the results of
calculations in our $Fe_8$ cluster "as it is" (with no longitudinal
field compensation). To obtain the two lower curves for both $W_D$ and
$E_o$, the corresponding external longitudinal field was applied
for each value of the external transverse field to shift a position
of the longitudinal fields distribution back to zero (the longitudinal
field compensated sample). The shift in our $Mn_{12}$ sample is small.

\begin{figure}[ht]
\centering
\vspace{-2.5cm}
\hspace{-0.3cm}
\includegraphics[scale=0.38]{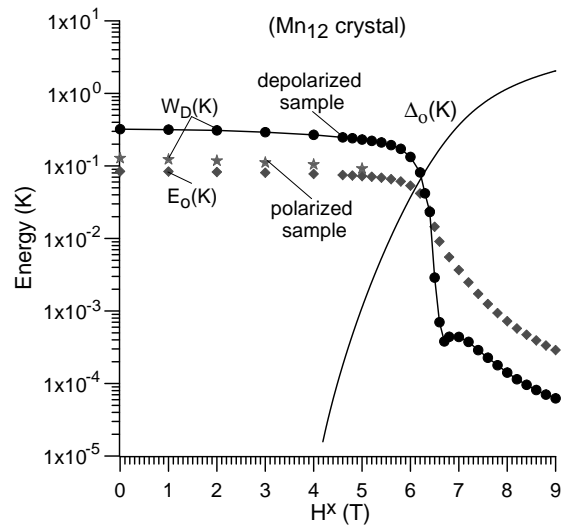}
\vspace{-1.5cm}
\caption{$Mn_{12}$ crystal (cluster of $50^3$ $Mn_{12}$ molecules;
no "faster relaxing species" \cite{DeltaDistr,WWEPL}). The curves are:
{\it solid line with circles} - $W_D(H^x)$ for "depolarized" sample;
{\it stars} - $W_D(H^x)$ for polarized sample; {\it diamonds} - $E_o(H^x)$;
{\it solid line} - $\Delta_o(H^x)$.}
\label{fig:DE_Fig2}
\end{figure}

\begin{figure}[ht]
\centering
\vspace{-2.0cm}
\hspace{-0.28cm}
\includegraphics[scale=0.38]{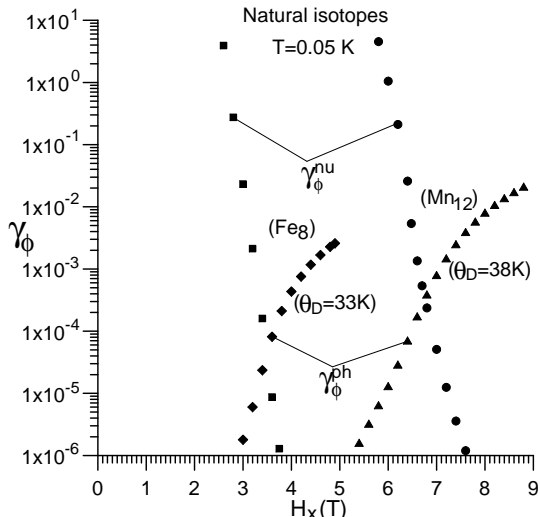}
\vspace{-2.0cm}
\caption{The transverse magnetic field behavior of the dimensionless
decoherence rates $\gamma^{nu}_{\phi}$ (shown by {\it squares} and
{\it circles}) and $\gamma^{ph}_{\phi}$ ({\it diamonds} and {\it triangles})
at azimuthal angle $\varphi=0$ (i.e., along the $x$-axis) for $Fe_8$
({\it squares} and {\it diamonds}) and $Mn_{12}$ ({\it circles} and
{\it triangles}) systems at $T=0.05 \; K$. The results are presented for
the natural isotopes ($Fe^{56}, \; H^1, \; Br^{79}, \; N^{14}, \; C^{12}$
and $O^{16}$ species for $Fe_8$ molecule and $Mn^{55}, \; H^1, \; C^{12}$
and $O^{16}$ species for $Mn_{12}$ molecule). }
\label{fig:DE_Fig3}
\end{figure}

\emph{\textbf{(5) NPC-window.}} ~ When studying the spin dynamics in the
NPC-window, one needs to know the region of the field where this window is
situated. This window can be rather narrow and for our examples of the $Fe_8$
and the $Mn_{12}$ systems this can be seen in Fig.\ref{fig:DE_Fig3}. Since we
are not going to discuss here a coherence optimization strategy, \cite{STPRB}
in this Figure we present the transverse field behavior of the dimensionless
decoherence rates $\gamma^{nu}_{\phi}$ and $\gamma^{ph}_{\phi}$,
Eq.(\ref{PH_NU}), for external field along the $x$ axis only and for molecules
containing only natural isotopes. \cite{NatIsot} (Note that both rates are
almost insensitive to changes in the populations of states $|\Uparrow \rangle$
and $|\Downarrow \rangle$.) The small oscillation energy $\Omega_o$ is $\sim E_g$
and, like $E_g(H^{\perp})$, slowly decreases with $H^{\perp}$. In zero field
$\Omega_o =2 S C_{\perp}(DE)^{1/2}$ ($C_{\perp} \approx 1.56$) for $Fe_8$ and
$\Omega_o \sim 2 S D$ for $Mn_{12}$. \cite{TB} The Debye energy $\Theta_D$
for $Fe_8$ and $Mn_{12}$ is known experimentally. \cite{Debye_MnFe}

\vspace{-3mm}

\section{Multimolecular processes in the limit $\Delta_o >> \{ W_D, E_o,
\hbar / \tau_{np} \}$}
\label{sec:mmproc}

If in the NPC-window the half-width $W_D$ of the dipolar bias
distribution is larger than $\Delta_o$, any spin dynamics is, in
general, incoherent. Outside of this window, independently of the
ratio $\Delta_o / W_D$, the spin dynamics is also incoherent. In
this Section we study the multimolecular correlations induced by
the dipole-dipole interactions between molecules in states
$|\Uparrow \rangle$ and $|\Downarrow \rangle$ in the {\it NPC}-window.
We assume that in the region of fields of our interest $\Delta_o >>
\{ W_D, E_o, \hbar / \tau_{np} \}$ (like, for example, in the region
$3.5 < H^x < 4.2 \; T$ for $Fe_8$ and in the region $6.7 < H^x < 7.4\; T$
for $Mn_{12}$, see Fig.\ref{fig:DE_Fig3}).

\subsection{One-pair processes}
\label{subsec:2M}

At very low temperatures, when only two lowest states of each molecule are
occupied, in the limit $\Delta_o >> \{W_D, E_o, \hbar / \tau_{np} \}$ we
work in the representation (\ref{TLS.2}) $\{ |\Uparrow \rangle,
|\Downarrow \rangle \}$ of the Hamiltonian (\ref{TLS.1}) (Appendix
\ref{sub:apA}). Consider one pair of interacting molecules in the sample
(Appendix \ref{sec:2TLS}). Such a pair is described by the Hamiltonian
(\ref{HM2}) ($\xi_i$ is the bias energy; in general, it is the
time-dependent parameter) and can be found in four states:
$|\Uparrow_1 \Uparrow_2 \rangle; \; |\Uparrow_1 \Downarrow_2 \rangle; \;
|\Downarrow_1 \Uparrow_2 \rangle; \; |\Downarrow_1 \Downarrow_2 \rangle$.
In the limit $\Delta_o >> |V^{\alpha \beta}_{dd}|$ two central "flip-flop"
states $|\Uparrow_1 \Downarrow_2 \rangle$ and $|\Downarrow_1 \Uparrow_2
\rangle$, linked by the effective tunneling matrix element
\begin{equation}
\Delta_{ff} \sim |V^{zz}_{dd}(R)| { \Delta_1 \Delta_2 \over \varepsilon_1
\varepsilon_2 }
\label{D_eff}
\end{equation}
(only the largest term in $B(\vec{R})$ Eq.(\ref{HM5}) is shown - its
form is similar to that known from the theory of dielectric glasses
\cite{Laikhtm,BurKag}), can be considered as an effective two-level
system with the asymmetry
\begin{equation}
\xi_{ff} \approx |\varepsilon_1 - \varepsilon_2| \sim |\xi^2_1 - \xi^2_2|
/ 2 \Delta_o << \Delta_o
\label{XI_eff}
\end{equation}
(recall that all $\Delta_i$ are supposed to be the same). Note that
all $V^{\alpha \beta}_{dd} (R)$ are {\it independent} of the external
field.

Two other states are separated from the two flip-flop states by the
energy gaps $> \Delta_o$ and in the region of fields where $\Delta_o >
|V^{\alpha \beta}_{dd}|$, their effect on the flip-flop transitions is
small. If $\Delta_{ff} > \xi_{ff}$, two flip-flop states are {\it in
resonance} and the amplitude of oscillations (with frequency $E_{ff}
\sim (\xi^2_{ff} + \Delta^2_{ff})^{1/2}$, (\ref{FF.2})) between them is
$\approx 1$. At the same time, transitions between other states are not in
resonance (they are accompanied by the energy change $> \Delta_o$) and
their amplitude is $\lesssim (V^{\alpha \beta}_{dd})^2 \xi^2_i / \Delta^4_i$.

In the field region where $\Delta_o \lesssim |V^{\alpha \beta}_{dd}|$
it is more convenient to solve the problem for the dynamics of a pair
of interacting spins in the basis set (\ref{TLS.3}). \cite{DubSt}
However, in this limit it can be rather difficult (or even impossible)
to observe coherent dynamics in an ensemble of spins. First of all, in
this limit $W_D$ also can be $\gtrsim \Delta_o$ (like in $Fe_8$ and
$Mn_{12}$). Moreover, the variety of different collective processes
leads to additional phase randomness and, consequently, to suppression
of coherence.

In what follows we suppose to work only in the part of the NPC-window
where $\Delta_o > |V^{\alpha \beta}_{dd}|$ and the probability to observe
coherent spin dynamics is larger.

At low temperatures ($k_B T < \Delta_o$) a number of the molecules in the
excited state $|\Downarrow \rangle$ with energy $+ \varepsilon$ can be
estimated as $N_{ex}(T) \sim N_o e^{-\varepsilon / k_B T} /(e^{-\varepsilon
/ k_B T} + e^{\varepsilon / k_B T}) \sim N_o e^{-2 \Delta_o / k_B T}$ ($N_o$
is the total number of molecules) and is small compared to the number of
the ground state molecules. These excited molecules are uniformly distributed
over the sample, and each of them is surrounded by the $\sim N_o / N_{ex}$
ground state molecules. The excited molecule can be in a "{\it flip-flop
resonance}" with the ground state molecule only if $\xi_{ff} \lesssim
\Delta_{ff}$. The time needed for the flip-flop transition to happen is $\sim
\hbar / \Delta_{ff} \sim O(R^3)$ and the fastest transitions are expected to
be between the nearest-neighbor molecules. For the effective two-level systems
composed of the two nearest-neighbor molecules we introduce the corresponding
effective tunneling matrix element $\Delta^{nn}_{ff}$ and the asymmetry
$\xi^{nn}_{ff}$.

In a simple cubic lattice each excited molecule can make a flip-flop
transition with any of its six nearest-neighbor ground state molecules
with the same probabilities. In a generic lattice these probabilities
can be different since $\Delta^{nn}_{ff}$ depends on the lattice structure.
The average over three crystallographic axes value of the $\Delta^{nn}_{ff}$
is $\sim W_D(H^{\perp}=0)$ for polarized sample, see Figs.\ref{fig:DE_Fig1}
- \ref{fig:DE_Fig2} ($W_D(0)$ for polarized sample is $\sim E_D / V^{(1)}_o$,
where $V^{(1)}_o$ is the volume per one molecule).

If $\Delta^{nn}_{ff} > \xi^{nn}_{ff}$ for the overwhelming majority of
the nearest-neighbor molecules (this issue is discussed in Section
\ref{subs:DEtoXI}), it is unlikely that at low temperatures any resonant
pair of the nearest-neighbor molecules (say, $i$-th and $j$-th molecules)
will remain in resonance for a long time. Instead, since the total probability
for the excited molecule (either $i$-th, or $j$-th, as a result of
oscillations (\ref{FF.1})) to create a resonance with one of the other
five nearest-neighbor molecules is larger than the probability to remain
in resonance with the same molecule all the time $t \sim \tau_{np}$, the
fastest flip-flop transitions can "propagate" through the crystal involving
more and more new molecules. Of course, not only the nearest-neighbor
molecules can be involved, but also the "lengthy" pairs (with $\Delta_{ff}(R)
< \Delta^{nn}_{ff}$). However, flip-flop transitions between the
nearest-neighbor molecules are faster.

In what follows, for brevity, these "mobile" (or "potentially
mobile") flip-flop transitions between the states $|\Uparrow \rangle$ and
$|\Downarrow \rangle$ in the {\it nearest-neighbor} molecules will be
called "{\it flipons}" (a kind of magnon). The number of flipons is
determined by the number of excited molecules $N_{ex}(T)$. In a generic
lattice $\Delta^{nn}_{ff}$ can be different along different crystallographic
axes. However, if flipon moves along one axis and for this axis
$\{ (\Delta^{nn}_{ff})_i \}$ are large in comparison with
$\{ (\xi^{nn}_{ff})_i \}$, such a movement is, in some sense, "coherent"
since flipon leaves site $i$ only because of equal probabilities for the
excited spin to create a resonance with both of its nearest neighbors
along this axis.

It is worth mentioning that, if there is a whole distribution of $\Delta_o$
(say, if there are "faster relaxing species" \cite{DeltaDistr,WWEPL}) in a sample,
the fraction of resonant flip-flop molecules decreases. This is because in such a
sample for some fraction of pairs the asymmetry $\xi_{ff}$ can be $\sim \Delta_o$
(and $\Delta_o$ increases with $H^{\perp}$). These impurities can essentially limit
(or even completely block) the motion of flipons.

\subsection{Multi-pair processes}
\label{subsec:MPP}

On average, two nearest-neighbor {\it excited} molecules are separated
by the distance
\begin{equation}
R_{ex}(T) \sim (V^{(1)}_o N_o / N_{ex})^{1/3}.
\label{R_EX}
\end{equation}
At $k_B T
<< \Delta_o$, $R_{ex}(T)$ is large compared to $ \widetilde{a} \equiv
(V^{(1)}_o)^{1/3}$ (in a cubic lattice $\widetilde{a} \equiv a$; $a$ is the
lattice constant). Consider two pairs of the {\it resonant nearest-neighbor}
molecules and let the distance between these pairs be $R$. This group of
four molecules contains two excited molecules (with energies $\varepsilon_2,
\; \varepsilon^{'}_2$) and two ground state molecules (with energies
$\varepsilon_1, \; \varepsilon^{'}_1$). In the limit $\Delta_o >> \{W_D, E_o,
\hbar / \tau_{np} \}$ and $\Delta_o > |V^{\alpha \beta}_{dd}|$ both these
resonant pairs experience mainly the flip-flop transitions.

If $R >> \widetilde{a}$, the strength $|V^{\alpha \beta}_{dd}(R)|$ of
the interactions between molecules belonging to different pairs is $<
\Delta^{nn}_{ff}$. Then, using the same arguments as for one pair
of molecules, in the case of two {\it resonant} pairs we can also
consider only corresponding collective "flip-flop" transitions between
the eigenstates of each resonant pair. The effective tunneling
matrix elements connecting these collective flip-flop states (separated
from all other states by the energy gaps $> \Delta^{nn}_{ff}$) is
\begin{equation}
\Delta^{(2)}_{ff}(\vec{R}) \sim |V^{zz}_{dd}(\vec{R})|
{ \Delta^{nn}_{ff} \Delta^{'nn}_{ff} \over E^{nn}_{ff} E^{'nn}_{ff} },
\label{D_eff_2}
\end{equation}
where $E^{nn}_{ff} \sim ((\xi^{nn}_{ff})^2 + (\Delta^{nn}_{ff})^2)^{1/2}$.
Similarly to the case of one pair of molecules, these collective flip-flop
states can also be considered as an effective two-level system with the
asymmetry
\begin{equation}
\xi^{(2)}_{ff} \sim |E^{nn}_{ff} - E^{'nn}_{ff}|.
\label{XI_eff_2}
\end{equation}
Note that we deliberately consider two pairs of nearest-neighbor
molecules since transitions between such molecules are faster, and in the
limit of our interest the probability to find them in resonance is larger.

The effective matrix element $\Delta^{(2)}_{ff}$ describes the flip-flop
transitions between the eigenstates of two resonant pairs (i.e.,
of two effective TLS). If $\Delta^{(2)}_{ff} > \xi^{(2)}_{ff}$,
two resonant pairs are in resonance with {\it each other}. The flip-flop
transitions between the eigenstates $|\Uparrow \rangle$ and
$|\Downarrow \rangle$ of each molecule inside of one resonant pair
are described by the matrix element $\Delta^{nn}_{ff}$. Then, since
$\Delta^{nn}_{ff} > \Delta^{(2)}_{ff}$, the frequency of oscillations
between states $|\Uparrow \rangle$ and $|\Downarrow \rangle$ of each
molecule in such a resonant group of four molecules is $\sim
\Delta^{nn}_{ff}$, but the group correlation time is $\sim \hbar
/ \Delta^{(2)}_{ff}$.

In a generic lattice, if the nearest-neighbor molecules in two pairs are
located along different axes, the asymmetry $\xi^{(2)}_{ff}$ can be $\sim
\Delta^{nn}_{ff} >> \Delta^{(2)}_{ff}$ and such two pairs can be out of
resonance. However, if molecules in both pairs are located along the same
axis (with a lattice constant $a$) and if $\Delta^{nn}_{ff} >> \xi^{nn}_{ff}$,
the asymmetry is
\begin{equation}
\xi^{(2)}_{ff} \sim |V^{zz}_{dd}(a)| f(\xi_i / \Delta_o), \;\;\;
f(\xi_i / \Delta_o) \sim O(\xi^2_i / \Delta^2_o),
\label{XI_eff2N}
\end{equation}
where the average value of $f(\xi_i/\Delta_o)$ can be estimated roughly
as $\sim W^2_D / \Delta^2_o$. Then, for the average asymmetry one gets
$\widetilde{\xi}^{(2)}_{ff} \sim |V^{zz}_{dd}(a)| (W_D / \Delta_o)^2 <<
|V^{zz}_{dd}(a)|$ and for such pairs the condition $\Delta^{(2)}_{ff}
\gtrsim \xi^{(2)}_{ff}$ can be, in principle, fulfilled (actually, the
difference of two mean energies $A$ and $A^{'}$ (\ref{HM5}) also contributes
to $\xi^{(2)}_{ff}$ and this gives a similar effect). The term neglected in
(\ref{XI_eff2N}) is $\sim (\Delta^{nn}_{ff} / W_D(H^{\perp}))^2$ times
smaller than the retained one - in the field region of our interest
in most systems $\Delta^{nn}_{ff}(\widetilde{a}) / W_D(H^{\perp}) >> 1$.

Note that for resonant pairs composed of the flip-flop molecules with
$\Delta_{ff}(R) << \Delta^{nn}_{ff}$, the glass-like scenario \cite{BurKag}
can be realized. In this case two pairs can be in resonance only if
$\Delta^{(2)}_{ff} \sim \Delta_{ff}$ (for most of such pairs $\xi^{(2)}_{ff}
\sim \Delta_{ff}$).

Knowing the sample average value of the asymmetry $\widetilde{\xi}^{(2)}_{ff}$,
from the requirement $\Delta^{(2)}_{ff}(\vec{R}) > \xi^{(2)}_{ff}$ one
can estimate the average "resonant" distance between two pairs:
\begin{equation}
R^{(2)}_{res}(H^{\perp}) \sim [ V^{(1)}_o \widetilde{V}_{dd} /
\widetilde{\xi}^{(2)}_{ff}(H^{\perp}) ]^{1/3},
\label{R_r}
\end{equation}
where \cite{NN_Str} $\widetilde{V}_{dd} \sim E_D / V^{(1)}_o$. If $R <
R^{(2)}_{res}$, two pairs could be, in principle, in resonance with each
other. However, even if $\widetilde{\xi}^{(2)}_{ff} \to 0$, not any two
pairs are in resonance since if $R > R_{ph}$, where
\begin{equation}
R_{ph}(T,H^{\perp}) \sim [V^{(1)}_o \widetilde{V}_{dd}
\tau^{ph}_{\phi}(T,H^{\perp}) / \hbar]^{1/3},
\label{R_PH}
\end{equation}
the two-pairs correlation time is longer than incoherent phonon-assisted
transitions in each molecule. Only the pairs satisfying the condition
$R < R_m = \min \{ R^{(2)}_{res}, R_{ph} \}$, can be in resonance. Thus,
if $R_{ex}(T,H^{\perp}) < R_m$, most of the closest {\it pairs} of resonant
molecules are able to come into resonance with each other. This happens
at temperatures $T > T_M$,
\begin{equation}
T_M = \max \{ T^{(2)}_{res}, T_{ph} \},
\label{T_M}
\end{equation}
with $T^{(2)}_{res}$ and $T_{ph}$ given by ($k_B T < \Delta_o$):
\begin{equation}
k_B T^{(2)}_{res}(H^{\perp}) \sim 2 \Delta_o / \ln [\widetilde{V}_{dd} /
\widetilde{\xi}^{(2)}_{ff}],
\label{T_res}
\end{equation}
\begin{equation}
k_B T_{ph}(H^{\perp}) \sim 2 \Delta_o / \ln [\Theta^4_D \widetilde{V}_{dd} /
S^2 \Omega^2_o \Delta^3_o ].
\label{T_ph}
\end{equation}
If at these conditions $\tau_{np} > t_c$, where
\begin{equation}
t_c \sim \hbar / |V^{zz}_{dd}(R_{ex})| \sim (\hbar / \widetilde{V}_{dd})
(R^3_{ex} / V^{(1)}_o),
\label{t_corr}
\end{equation}
at $t \gtrsim t_c$ the whole hierarchy of (more or less) correlated
flip-flop clusters of the increasing "size" $n$ (the number of involved
resonant flip-flop pairs) can, in principle, appear. The time $t_c$
estimates the cluster correlation time.

Note, however, that if $\Delta^{nn}_{ff} > \xi^{nn}_{ff}$ for most of
the nearest-neighbor molecules, instead of interactions between fixed
pairs of resonant molecules, in the limit of our interest we have a
set of {\it flipons} moving through the sample and interacting with
each-other. At $T < T_M$ they participate in the collective processes
very rarely (interactions can be neglected). When temperature increases,
the number of flipons also increases and collective processes become
more frequent. At $T > T_M$ correlations between flipons, in principle,
may still lead to the creation of correlated clusters. However, due to
various decorrelation (dephasing) processes, these clusters can be
destroyed rapidly (or they will not be able to appear at all). Such
dephasing processes will be considered in Section \ref{subsec:dephas}.

\subsection{Decorrelation}
\label{subsec:dephas}

\emph{\textbf{(1) Flipon motion.}} ~ If flipons are delocalized, the
effective tunneling matrix element $\Delta^{(2)}_{ff}$ changes with the
distance between flipons resulting in the suppression of correlations in
clusters (if they appear).

Suppose that at $t=0$ there is a correlated cluster ($T_M < T < \Delta_o
/ k_B$) composed of nearly equidistant flipons (with distance $\approx
R_{ex}$). If $\Delta^{nn}_{ff} >> \xi^{nn}_{ff}$ and if the flipons move
along the same axis, correlations between them will not necessarily be
destroyed immediately after the first "jump". Of course, if at $t > 0$ the
flipons start to move along different axes, in a generic lattice any
correlations can be destroyed almost immediately (i.e., at $t \gtrsim \hbar
/ \Delta^{nn}_{ff}$) since in this case $\xi^{(2)}_{ff}$ can become $\sim
\Delta^{nn}_{ff} >> \Delta^{(2)}_{ff}$. Note, however, that the flipons have
larger probability to move along the axis with shortest lattice constant.
Then, if we consider only a {\it quasi}-$1d$ motion of flipons (i.e., along
the same axis), we can estimate the {\it longest} "motional" dephasing time
$\tau^m_d$. Comparison of this time with the cluster correlation time $t_c$
($t_c < \tau_{np}$) shows whether the correlated cluster with the average
distances $R_{ex}$ between the nearest-neighbor flipons can appear.

For the sake of simplicity, we approximate the flipon centers of mass
motion by the {\it discrete} "random walks" model (Appendix
\ref{sub:Rand_W}). At $t>0$ the distances $R(t)$ between the
nearest-neighbor flipons in the whole cluster become distributed around
$R_{ex}$ with nonzero half-width $\delta R(t)$. Thus, instead
of a single "line" $\Delta^{(2)}_{ff}(R_{ex})$ one also gets a whole
distribution of values $\Delta^{(2)}_{ff}(R)$ with nonzero mean-square
deviation $\delta \Delta^{(2)}(t) = \langle (\Delta^{(2)}_{ff})^2
- \langle \Delta^{(2)}_{ff} \rangle^2 \rangle^{1/2}$. Knowing
$\delta \Delta^{(2)}(t)$, the motional dephasing time can be obtained
from the condition
\begin{equation}
\tau^m_d = t_f N^m_d; \;\;
\sum^{N^m_d}_{N=0} t_f \delta \Delta^{(2)}(N) \sim \hbar; \;\;
t_f = \hbar / \Delta^{nn}_{ff}
\label{t_mdef}
\end{equation}
for $N = t / t_f$ (or from the condition $\int^{\tau^m_d}_0 d t \;
\delta \Delta^{(2)}(t) \sim \hbar$ at large values of $N$). Obviously,
the correlations in the whole cluster will be destroyed together with the
destruction of resonances between the nearest-neighbor flipons. Since
in each pair both flipons can move, for $\langle \Delta^{(2)}_{ff}
\rangle$ we have
\begin{equation}
\sim \sum^N_{\widetilde{r}_1, \widetilde{r}_2 = -N}
|V^{zz}_{dd}(R_{ex} + (\widetilde{r}_1 - \widetilde{r}_2) \widetilde{a})|
P_N(\widetilde{r}_1) P_N(\widetilde{r}_2)
\label{D_sr}
\end{equation}
with the condition $(R_{ex} + (\widetilde{r}_1 - \widetilde{r}_2)
\widetilde{a}) / R_{ex} \ge \eta(T)$. Here $\eta(T)$ is the dimensionless
(in units of $R_{ex}(T)$) minimally possible distance between the centers
of mass of two flipons. Each distribution $P_N(\widetilde{r}_i)$ gives the
probability to find the $i$-th flipon at the distance $\widetilde{r}_i
\widetilde{a}$ ($\widetilde{r}_i < N$) from its $t=0$ position after total
$N$ steps (Appendix \ref{sub:Rand_W}). \cite{2FDF}

The solution of Eq.(\ref{t_mdef}) depends on $\eta$ and $\rho$
(Eq.(\ref{FDC})) and can be found numerically. For $p=q=s=1/3$
(Eq.(\ref{Polyn1})) and $\rho = 2/3$ we get
\begin{equation}
\tau^m_d = N^m_d t_f \equiv 3 \lambda(\eta) [R_{ex} / \widetilde{a}]^2 t_f
\sim 3 \lambda(\eta) [\widetilde{a} / R_{ex}] t_c.
\label{tau_md}
\end{equation}
If $\eta = 2 \widetilde{a} / R_{ex}$, for $0.05 \le \eta \le 2/3$ we get
$\lambda \equiv \lambda_2 \approx 1.6 \eta^2 + 0.35 \eta + 0.045$. If
$\eta = \widetilde{a} / R_{ex}$, for $0.05 \le \eta < 1/2$ we get $\lambda
\equiv \lambda_1 \approx 0.34 \eta^2 + 0.2 \eta + 0.03$. Note that the
configurations of the nearly equidistant flipons with $R_{ex} = \widetilde{a}$
do not exist, in contrast to those with $R_{ex} = 2 \widetilde{a}$.
However, if flipons move along the same axis, but in the nearest-neighbor
rows, the centers of mass of some flipons can be separated by the distance
$\widetilde{a}$. To take this effect into account, one can use $\lambda =
(\lambda_2 + \lambda_1) / 2$ for estimations.

The answer for $\tau^m_d$ can be found in the equivalent dimensionless
form $D_f \tau^m_d / R^2_{ex} = \widetilde{\lambda}(\eta,\rho)$, where
$D_f$ is the flipon effective diffusion coefficient (\ref{FDC}) and
at $\rho = 2/3$, $\widetilde{\lambda}(\eta,\rho) \equiv \lambda(\eta)$
from Eq.(\ref{tau_md}). The $\rho$-dependence of $\tau^m_d$ is roughly
$\sim \rho^{-1/3}$. Then, $\tau^m_d / t_c$ can become larger
either (i) at $T \to \Delta_o / k_B$, when flipons are in their dense
phase and essentially localized ($\delta R(t) \to 0$); or (ii) if $s =
1 - p - q \to 1$, $\rho \to 0$ and flipons are almost immobile
even at $T < \Delta_o / k_B$. The latter can be, in principle, realized
in a sample with impurities.

If $\tau^m_d(T) << t_c(T)$, the creation of a correlated cluster at the
average distance $R_{ex}(T)$ is virtually impossible. Solving either the
equation $N^m_d(R_c) t_f = t_c(R_c)$, or the equation
\begin{equation}
D_f t_c(R_c) / R^2_c = \widetilde{\lambda}(\eta_c,\rho),
\label{R_c}
\end{equation}
one finds the average distance $R_c$ and the temperature
\begin{equation}
k_B T_c \sim 2 \Delta_o(H^{\perp}) / \ln [(R_c / \widetilde{a})^3 - 1]
\label{T_c}
\end{equation}
at which $\tau^m_d \sim t_c$ and cluster can appear. For $\rho = 2/3$
and $\eta_c = 2 \widetilde{a} / R_{c}$ we get $R_c \sim 3 \widetilde{a}$
and $k_B T_c \sim 0.6 \Delta_o(H^{\perp})$. For $\rho = 2/3$ and $\eta_c
= \widetilde{a} / R_c$ we get $R_c \sim 2 \widetilde{a}$ and $T_{c} \sim
\Delta_o(H^{\perp})$. These estimations shows that, if the scenario with
$p \approx q \approx s \approx 1/3$ is realized, $\tau^m_d(T) < t_c(T)$
almost everywhere except the flipons dense phase at $T \to \Delta_o / k_B$,
where $\tau^m_d(T) \sim t_c(T)$ and where $t_c(T)$ decreases itself (as well
as $\tau_{np}(T)$, see Eq.\ref{PH_NU}). If, in contrast, $s \to 1$ and
$\rho \to 0$, $\tau^m_d >> t_c$ and correlated clusters can appear even at
$T_M < T < T_c$.

\emph{\textbf{(2) "Spectral diffusion".}} ~ The above described picture
is valid only if $\Delta^{nn}_{ff} > \xi^{nn}_{ff}$ for most of the
nearest-neighbor molecules. In the opposite limit, at $T > T_M$ the
correlated clusters will be composed of almost immobile "lengthy"
flip-flop pairs with $\Delta_{ff} (R > \widetilde{a}) < \Delta^{nn}_{ff}$,
satisfying the condition $\Delta_{ff} \sim \Delta^{(2)}_{ff}$. This
scenario is very similar to that in dielectric glasses \cite{BurKag}
and the cluster dephasing time at $t < \tau_{np}$ will be determined
by the process similar to the "spectral diffusion" in glasses.
\cite{Laikhtm,KlBlPh} The change of states of fixed effective TLS
results in the bias fluctuations and, consequently, in the dephasing.
In this limit ($\Delta^{nn}_{ff} < \xi^{nn}_{ff}$) the cluster "spectral
diffusion" dephasing time $\tau^s_d$ is $ \sim t_c$ since the asymmetry
$\xi^{(2)}_{ff}$ for most groups of two resonant pairs is $\sim \Delta_{ff}$.

In the limit $\Delta^{nn}_{ff} > \xi^{nn}_{ff}$ the spectral
diffusion-like process can contribute as well. Instead of going deeply
into the details, here we only estimate the corresponding effects.
Consider, for simplicity, the case of immobile flipons.
The bias $(\xi^{nn}_{ff})_i$, acting on any $i$-th flipon, contains
contributions from all individual spins in the sample. When any $j$-th
flipon makes a transition, the {\it change} of the bias, acting on
$i$-th flipon, is $\delta^f_{ij}(R_{ij}) \sim V^{zz}_{dd}(R_{ij}) \;
(\xi^{nn}_{ff})_i (\xi^{nn}_{ff})_j / (E^{nn}_{ff})_i (E^{nn}_{ff})_j$.
(Here $(\xi^{nn}_{ff})_{i,j}$ can be both positive and negative and the
term $V^{zx}_{dd} \tau^z_i \tau^x_j$ does not change its sign when
$j$-th flipon makes transition.) Then, if $N_t$ flipons make a
transition ($\max(N_t) = N_{ex}$), the total change of the bias acting
on $i$-th flipons is $\delta^f_i = \sum^{N_t}_{j=1} \delta^f_{ij}(R_{ij})$.

Depending on the degree of "polarization" $M_t = (N_{gs} - N_{es}) /
(N_{gs} + N_{es})$ of the group of $N_t$ flipons ($N_{gs}$
and $N_{es}$ are the numbers of flipons in their ground and excited
states), the total bias change $|\delta^f_i|$ can vary
roughly from $\widetilde{\delta}^f_i \sim |V^{zz}_{dd}(R_{ex})| \;
[(\xi^{nn}_{ff})_i / (\Delta^{nn}_{ff})_i]
[\widetilde{\xi}^{nn}_{ff} / \widetilde{\Delta}^{nn}_{ff}]$ to $\sim
\widetilde{\delta}^f_i \ln (N_t)$ ($\widetilde{\Delta}^{nn}_{ff}$ is
the sample average value of $(\Delta^{nn}_{ff})_j$ and
$\widetilde{\xi}^{nn}_{ff}$ is the sample average absolute value of
$(\xi^{nn}_{ff})_j$). Then the shortest dephasing time $\tau^s_d$ is
$\sim t_f$. However, the dipole-dipole interaction changes its sign
with the direction of $\vec{R}_{ij}$ and, on average, in the case
$M_t \to 1$ the "surface" spins will mainly determine the maximum
value of the bias change $|\delta^f_i|$. For spins (flipons) in the
bulk this essentially reduces $|\delta^f_i|$ and increases
$\tau^s_d$.

If $M_t \to 0$, simultaneous transitions of many flipons nearly
cancel the effect of each other resulting in $|\delta^f_i| \lesssim
\widetilde{\delta}^f_i << |V^{zz}_{dd}(R_{ex})| << \Delta^{nn}_{ff}$. In
this case the spectral diffusion mechanism cannot destroy the resonance
neither inside of the individual flipons, nor between them. Indeed, the
contribution from the $\xi^{nn}_{ff}$ to the asymmetry $\xi^{(2)}_{ff}$
in the limit $\Delta^{nn}_{ff} > \xi^{nn}_{ff}$ is $\sim (\xi^{nn}_{ff})^2
/ 2 \Delta^{nn}_{ff}$ and its change due to $\delta^f_i$ is $\sim \delta^f_i
\; (\xi^{nn}_{ff})^i / (\Delta^{nn}_{ff})^i$.

On average, in a temperature
equilibrated sample, it is plausible to assume $M_t \to 0$ and in the limit
$\Delta^{nn}_{ff} > \xi^{nn}_{ff}$ the spectral diffusion effect is much
weaker than the motional dephasing effect (broadening of the
$\Delta^{(2)}_{ff}$ distribution due to the flipons motion). This remains
valid also if the flipons are allowed to move. In this case the spectral
diffusion dephasing time $\tau^s_d$ is roughly $\sim
(\widetilde{\Delta}^{nn}_{ff} / \widetilde{\xi}^{nn}_{ff})^3$ times longer
than the motional dephasing time $\tau^m_d$ (as it was already noted, if
flipons move along different axis, in a generic lattice correlations can
be destroyed already at $t \sim t_f$).

\section{Discussion}
\label{sec:spindyn}

In this Section we discuss a temperature equilibrated sample
at $k_B T \lesssim \Delta_o$, where only the fraction $N_{ex}(T)$ of
molecules are in the excited states $|\Downarrow \rangle$. Thus, we are
not going to discuss the problem of the magnetization relaxation in the
limit $\Delta_o >> \{ W_D, E_o \}$. If, for example, the sample is
prepared at very high temperatures and then rapidly cooled down to low
temperatures, it will start to relax to its temperature equilibrated
state - this relaxation process will not be discussed here either.

If, in some system, in all the NPC-window $\Delta_o \lesssim
|V^{\alpha \beta}_{dd}|$, it can be very difficult (most probably,
impossible) to observe any coherent spin dynamics in an ensemble of
interacting spins (Section \ref{subsec:2M}). Here we consider only
the part of the NPC-window, where $\Delta_o > |V^{\alpha \beta}_{dd}|$.
In both the $Fe_8$ and the $Mn_{12}$
systems, which are used in this work to illustrate how the problem
can be analyzed, this field region is situated to the right of the
minimum of $\gamma^{nu}_{\phi} + \gamma^{ph}_{\phi}$ (see
Fig.\ref{fig:DE_Fig3}). The collective processes, having the largest
amplitude in this region of fields, are the flip-flop processes.

\subsection{The ratio $\widetilde{\Delta}^{nn}_{ff} /
\widetilde{\xi}^{nn}_{ff}$}
\label{subs:DEtoXI}

The fastest pair flip-flop processes are the processes between the
nearest-neighbor molecules. The average strength of the
nearest-neighbor dipole-dipole interactions and the average value
of the flipon effective tunneling matrix element
$\widetilde{\Delta}^{nn}_{ff}$ are $\sim \widetilde{V}_{dd}$
($\widetilde{V}_{dd} \sim 0.12 \; K$ for $Fe_8$ and $\sim 0.07 \; K$
for $Mn_{12}$). To find the average asymmetry $\widetilde{\xi}^{nn}_{ff}$,
we first calculate the distributions $P_{12}(\varepsilon,H^{\perp})$
of the $\varepsilon = \varepsilon_1 - \varepsilon_2$ ($\varepsilon_i
= (\Delta^2_i + \xi^2_i)^{1/2}$; all $\Delta_i$ depend on both the
external and the dipolar transverse fields and are obtained by the
exact diagonalization of (\ref{Hfe}) and (\ref{Hmn}), Appendix
\ref{sub:Calc_meth}), where $\varepsilon_{1,2}$ are for the
nearest-neighbor molecules only. Then, we get
$\widetilde{\xi}^{nn}_{ff} = (1/2) \int^{+ \varepsilon_o}_{- \varepsilon_o}
d \varepsilon \; |\varepsilon| P_{12}(\varepsilon,H^{\perp}) $
($\varepsilon_o$ is the maximum value of $\varepsilon$).

\begin{figure}[ht]
\centering
\vspace{-3.0cm}
\hspace{0.0cm}
\includegraphics[scale=0.4]{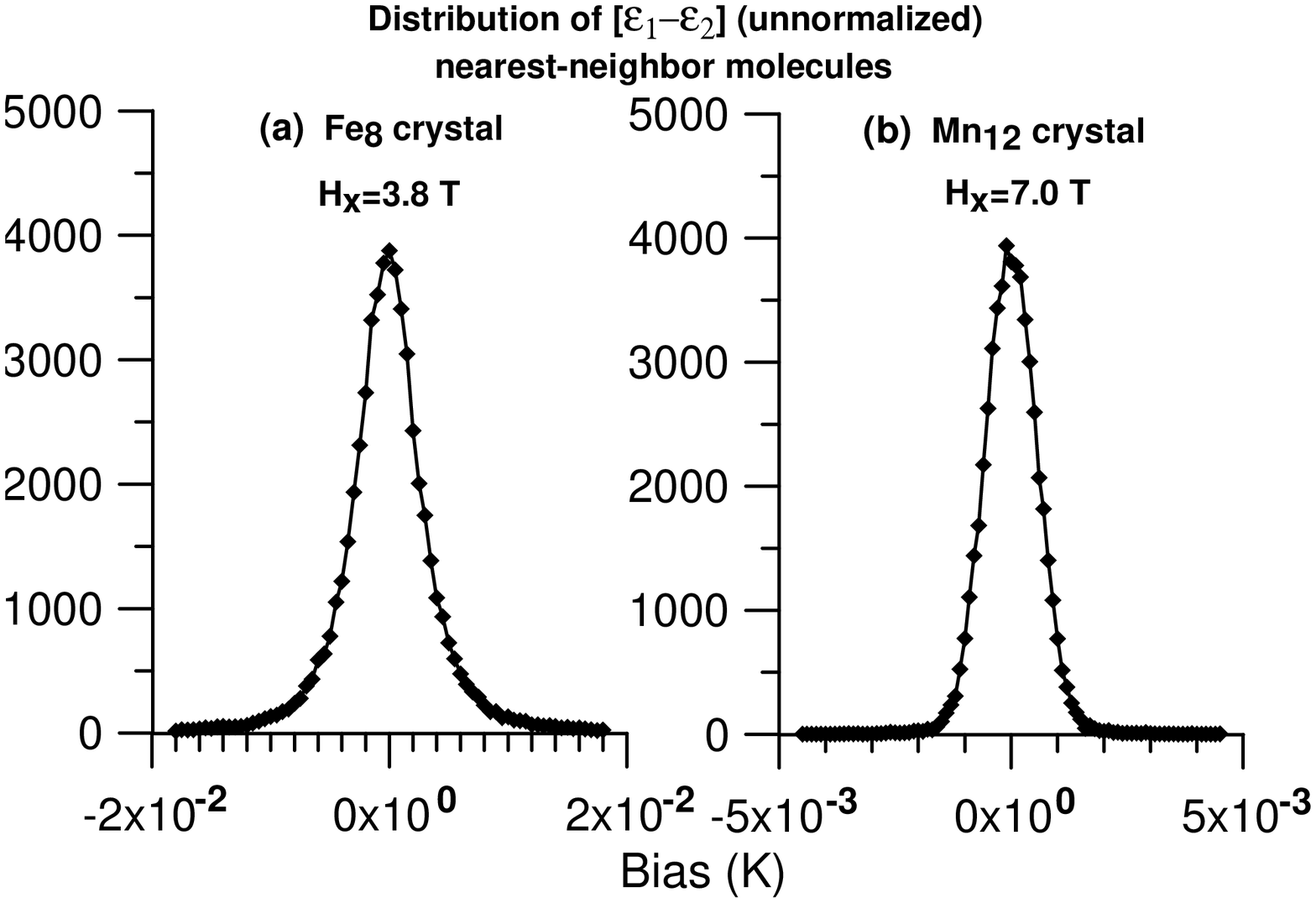}
\vspace{-3.5cm}
\caption{The distributions of
the $\varepsilon = \varepsilon_1 - \varepsilon_2$ for the nearest-neighbor
molecules: (a) at $H^x=3.8 \; T$ in the crystal of $Fe_{8}$ molecules (no
field compensation); (b) at $H^x=7.0 \; T$ in the crystal of $Mn_{12}$
molecules (no faster relaxing species). All cluster parameters are the same
as in Figs.\ref{fig:DE_Fig1} - \ref{fig:DE_Fig3}. Here $\varepsilon$ is in
Kelvins. The states $|\Uparrow \rangle$ and $|\Downarrow \rangle$ are
equipopulated. Both distributions are averaged over three crystallographic
axes.}
\label{fig:DE_Fig4}
\end{figure}

For example, two distributions (at $H^x = 3.8 \; T$ for $Fe_8$ and at
$H^x = 7.0 \; T$ for $Mn_{12}$) are presented in Fig.\ref{fig:DE_Fig4}.
For these distributions we get $\widetilde{\xi}^{nn}_{ff} \approx 1.5
\times 10^{-3} \; K$ for $Fe_8$ (with no field compensation) and
$\widetilde{\xi}^{nn}_{ff} \approx 2.3 \times 10^{-4} \; K$ for $Mn_{12}$
(the corresponding mean-square deviations $\delta \varepsilon = \langle
\varepsilon^2 - \langle \varepsilon \rangle^2 \rangle^{1/2}$ are
$\approx 4.7 \times 10^{-3} \; K$ and $\approx 6.4 \times 10^{-4} \; K$
respectively). For the field-compensated $Fe_8$ sample, the corresponding
values are about two times smaller.

To summarize the results, we note that under the assumption of the absence
of impurities with larger (or smaller) values of $\Delta_i$, in both systems
in the NPC-window the average asymmetry $\widetilde{\xi}^{nn}_{ff}$ (as well
as $\delta \varepsilon$) is small in comparison with
$\widetilde{\Delta}^{nn}_{ff}$ for most of the nearest-neighbor molecules. In
this case flipons can move in both systems.

Note, however, that in the $Mn_{12}$
crystal there are faster relaxing (minor) species \cite{DeltaDistr,WWEPL}
(about $5-10 \; \%$ of all molecules have lower potential barrier and larger
values of $\Delta_i$) - these species can change somehow both the
$\widetilde{\xi}^{nn}_{ff}$ and the $\delta \varepsilon$. In the NPC-window
for major species the spin dynamics of these minor species is incoherent
already. Since even the Hamiltonian for these minor species is still
unknown (or unpublished), we cannot describe their effect quantitatively.
Obviously, they can decrease the number of flipons and limit their motion.

\subsection{Collective spin dynamics}
\label{subs:2TR}

From Eqs.(\ref{T_res})-(\ref{T_ph}) in the $Fe_8$ crystal at $H^x=3.8 \; T$
($\Delta_o \approx 0.35 \; K$) one finds $T_M \sim 0.1 \; K$ for the field
compensated sample ($T_M \sim 0.11 \; K$ with no field compensation). The
corresponding temperature for the $Mn_{12}$ crystal at $H^x=7.0 \; T$
($\Delta_o \approx 0.36 \; K$) is $T_M \sim 0.13 \; K$. At $T < T_M$ the
average distance between flipons is large, and so the collective multi-pair
flip-flop processes are essentially "frozen" and correlated clusters of
flipons can not appear.

At $T_M < T \lesssim \Delta_o / k_B$ the collective multi-pairs processes
are unfrozen and, when the temperature increases, the whole hierarchy
of correlated clusters of increasing size $n$ (the number of involved
flipons, $\max \{ n \} = N_{ex}(T)$) can, in principle, appear. The
motion of flipons leads to the suppression of correlations. The spectral
diffusion effect in the limit $\Delta_o >> \{ W_D, E_o \}$ is rather
weak (Section \ref{subsec:dephas}) and the spectral diffusion dephasing
time $\tau^s_d$ is longer than the motional dephasing time $\tau^{m}_d$.
If at certain temperature the cluster dephasing time $\tau^{ms}_d(T) =
\min \{ \tau^m_d(T), \tau^s_d(T) \}$ is shorter than the cluster
correlation time $t_c(T)$, creation of the correlated clusters at
average distance $R_{ex}(T)$ is virtually impossible. Instead, only
short-living (with $\tau^{ms}_{d} < t_c < \tau_{nd}$) correlations
between molecules at distances $R < R_{ex}$ can be present.

If flipons can move, the correlated clusters at average distance $R_{ex}$
can appear only at temperatures $T \gtrsim T_c > T_M$ (Eqs.(\ref{T_M},
\ref{T_c})), when $\tau^{m}_d$ becomes longer than $t_c$ and
$R_{ex}(T)$ becomes shorter then $R_c$ (Eq.(\ref{R_c})). In $Fe_8$ (for
$H^x=3.8 \; T$) and in $Mn_{12}$ (for $H^x=7.0 \; T$) crystals this may
happen already at $T \sim 0.25 \; K$ (assuming $p=q=s=1/3$ and $\rho = 2/3$,
see Section \ref{subsec:dephas}). Note, however, that in the $Fe_8$ crystal
for example, where all lattice constants are different, only flipons
oriented (propagating) along the same crystallographic axis can create
correlated cluster. At the same time, flipons have larger probability to
propagate along the axis with the shortest inter-molecular distance (i.e.,
with larger $\Delta^{nn}_{ff}$). In this case a quasi-$1d$ motion of flipons
is more probable than a $3d$ one. Nevertheless, if at $t > 0$ flipons will
start to change their orientation, this process will speed up decorrelation.

If the cluster dephasing time $\tau^{ms}_d$ is $\sim t_c << \tau_{np}$
(Eq.\ref{taunp}), during the time interval $t = \tau_{np}$ the correlated
clusters can be created and destroyed (roughly) $\sim \tau_{np} / t_c$ times.
During the cluster life-time $\tau^{ms}_d$ all molecules, belonging to cluster,
can make $\sim \Delta^{nn}_{ff} t_c / \hbar$ coherent oscillations
(\ref{FF.1}). At $T \sim 0.25 \; K$ the correlated clusters can "reappear"
$\sim 50$ times in $Fe_8$ at $H^x = 3.8 \; T$ (for a cluster of flipons
oriented along the same axis) and $\sim 10$ times in $Mn_{12}$ at
$H^x = 7.0 \; T$. At these fields one may expect $\sim 30$ oscillations
(\ref{FF.1}) in both $Fe_8$ and $Mn_{12}$. All these estimations do not
take into account the coherence optimization strategy \cite{STPRB} and we
again assume $p=q=s=1/3$ and $\rho = 2/3$. Note also that the phonon
decoherence rate $\gamma^{ph}_{\phi}$ increases with temperature (see
Eq.\ref{PH_NU}), but at $k_B T < \Delta_o$ this increase is slow.

In the case of $Mn_{12}$, in the NPC-window for major species the spin
dynamics of molecules belonging to minor species is incoherent already.
Due to the difference in $\Delta_i$, the molecules of major and minor
species cannot create correlated flip-flop pairs between each other. This
results (i) in a decrease in the number of flipons; (ii) in partial
localization of flipons (depending on the fraction of impurities); and
(iii) in randomization of processes due to interactions between molecules
belonging to different species. The last effect gives rise to the
incoherent pair processes, leading to suppression of coherence.

The larger the concentration of impurities with incoherent internal
dynamics, the smaller the probability for correlated clusters to
appear. However, at $T > T_c$ the sample can become covered by
correlated clusters of the sizes smaller than the average distance
between impurities $R_{im}$. For example, in $Mn_{12}$ at $H^x = 7.0
\; T$ and $T = 0.25 \; K$ one gets $R_{ex} \sim 3 \widetilde{a}$. If
$R_{im} \gtrsim 10 \widetilde{a}$, at these values of field and
temperature the correlated clusters of the radius $\sim 3R_{ex}$
can, in principle, appear.

If $\Delta^{nn}_{ff} < \xi^{nn}_{ff}$ for most of the nearest-neighbor
molecules (i.e., no flipons), the correlated clusters composed of
{\it lengthy} flip-flop pairs with $\Delta_{ff}(R \sim R_{ex}) <<
\Delta^{nn}_{ff}$ can appear already at $T_M < T < T_c$. The
asymmetry $\xi^{(2)}_{ff}$ for two resonant pairs in such a cluster
is $\sim \Delta_{ff} \sim \Delta^{(2)}_{ff}$ and the cluster life-time
is $\sim t_c$. This means that all resonant pairs will be able to make
only one corelated flip-flop transition before correlations will be
suppressed (if $t_c << \tau_{np}$, such a cluster can reappear $\sim
\tau_{np} / t_c$ times).
If $\Delta^{nn}_{ff} > \xi^{nn}_{ff}$, but for some reason $\rho \to 0$
($s \to 1$), the correlated clusters of immobile flipons can appear.
Because of the weakening of the spectral diffusion effect in the limit
$\Delta_o >> \{ W_D, E_o\}$, the dephasing time for these clusters can
be limited only by $\tau_{np}$ and the number of oscillations (\ref{FF.1})
can be limited only by $Q_{\phi} \Delta^{nn}_{ff} / \Delta_o$.

\section{Summary}
\label{Summ}

In the present work the internal dynamics of a temperature-equilibrated
crystalline sample of the dipole-dipole interacting molecules with the central
spins $\vec{S}_i$ has been studied in the coherence window for nuclear spin and
phonon degrees of freedom.

At large external transverse magnetic fields the tunneling matrix
element $\Delta_o$ (Section \ref{sec:intr} and Appendix \ref{sub:apA})
increases, whereas both the half-width of the dipolar bias
distribution $W_D$ and the half-width of the hyperfine bias
distribution $E_o$ decrease (Section \ref{sec:model} and Appendix
\ref{sub:Calc_meth}).

At a certain value of the transverse field, the coherence window for phonon
and nuclear spin-mediated decoherence (the NPC-window) opens up (Sections
\ref{sec:intr} and \ref{sec:model}). Outside of the NPC-window the spin
dynamics is incoherent. If in the whole NPC-window the average strength of
the dipole-dipole interactions between the nearest-neighbor molecules
$\widetilde{V}_{dd}$ or $W_D$ are larger than $\Delta_o$, the spin
dynamics is also incoherent. In the opposite limit the coherent spin
dynamics is possible.

In the limit $\Delta_o > \widetilde{V}_{dd}$ and if the effective matrix
element $\Delta^{nn}_{ff}$, describing transitions between two flip-flop
states of the nearest-neighbor molecules, Eq.(\ref{D_eff}), is large
compared to the asymmetry $\xi^{nn}_{ff}$ of these two states,
Eq.(\ref{XI_eff}), the spin correlations between the nearest-neighbor
molecules lead to the creation of resonant flip-flop pairs (Section
\ref{subsec:2M} and Appendix \ref{sec:2TLS}). Such resonant pair
experiences oscillations (\ref{FF.1}) between states $|\Uparrow
\Downarrow \rangle$ and $|\Downarrow  \Uparrow \rangle$ of two
molecules with frequency $\sim \Delta^{nn}_{ff}$.

If $\Delta^{nn}_{ff} > \xi^{nn}_{ff}$ for the most pairs of the
nearest-neighbor molecules, the resonant flip-flop transitions can
"propagate" in the crystal, involving more and more new molecules (one
molecule in a pair remains in its ground state but another
nearest-neighbor molecule creates new resonance with the excited
molecule). This "mobile" magnon-like process (a spin-excitation)
between the states of two involved nearest-neighbor molecules in our
work is called "flipon" (Section \ref{subsec:2M}). The number of
flipons is limited by the number of excited molecules $N_{ex}(T)$.

At $T < T_M = \max \{ T^{(2)}_{res}, T_{ph} \}$,
Eqs.(\ref{T_res},\ref{T_ph}), the distances between flipons are long
and the correlations between them are unimportant. At $T > T_M$ the
correlations between flipons become crucial and at certain conditions
can lead to the creation of correlated clusters of flipons (Section
\ref{subsec:MPP}). Each cluster represents a correlated group of
molecules experiencing coherent oscillations (\ref{FF.1}) between
their lowest states $|\Uparrow \rangle$ and $|\Downarrow \rangle$.

The flipons motion and the spectral diffusion process result in
the suppression of correlations (Section \ref{subsec:dephas}). If
$\Delta^{nn}_{ff} >> \xi^{nn}_{ff}$ for most pairs of the
nearest-neighbor molecules (Section \ref{subs:DEtoXI}) and flipons
can move, the correlated cluster can appear only at $T \gtrsim T_c$,
Eq.(\ref{T_c}). The average inter-flipon distance in this case is
given by the solution of Eq.(\ref{R_c}) and the cluster dephasing
time (i.e., its life-time) $\tau^{ms}_d$ is $\sim t_c$ (Eq.(\ref{t_corr})).
Molecules that belong to the cluster can make $\sim \Delta^{nn}_{ff}
t_c / \hbar$ oscillations (\ref{FF.1}) before the correlations will
be suppressed. During the total phonon/nuclear spin coherence time
$\tau_{np}$, Eq.(\ref{taunp}), the coherent clusters can "reappear"
$\lesssim \tau_{np} / t_c$ times. At $T < T_c$ only random short-living
($t < t_c$) correlations within small groups of flipons, separated
by distances $R < R_{ex}(T)$, Eq.(\ref{R_EX}), can appear.

The smaller the effective flipon diffusion coefficient $D_f$,
Eq.(\ref{FDC}), the longer the cluster life-time $\tau^{ms}_d$.
If, for some reason, the flipons are localized ($D_f \to 0$), the
correlated clusters of immobile flipons can appear even at $T <
T_c$. Their life-time can be limited only by $\tau_{np}$, and the
number of oscillations (\ref{FF.1}) can be limited only by
$Q_{\phi} \Delta^{nn}_{ff} / \Delta_o$. If $\Delta^{nn}_{ff} <
\xi^{nn}_{ff}$, the correlated clusters of
"lengthy" resonant flip-flop pairs (molecules in these pairs are
separated by the distance $R \sim R_{ex}$) can appear also at
$T < T_c$. Their life-time is limited by $t_c$. On average, all
pairs in these clusters will be able to make only one flip-flop
transition before correlations will be suppressed. If $t_c <
\tau_{np}$, these clusters can reappear several times.

It is worth mentioning also that various systems allow the coherence
optimization strategy (orientation of external transverse field in a plane,
chemical replacement of isotopes, etc. \cite{STPRB}) to be applied to get
longer spin/phonon coherence time-interval $\tau_{np}$ or to shift the
coherence window down to lower values of the transverse field.

This concludes our study of the collective spin dynamics of a temperature
equilibrated sample in the coherence window for phonon and nuclear
spin-mediated decoherence. The presented analysis can be applied
to any crystalline nanomagnetic insulator composed of the central spin
$\vec{S}$ molecules, and can be useful in the search for magnetic systems
showing the spin coherence and collective phenomena. The analysis for the
induced dynamics (when system at $t=0$ is {\it prepared} in some initial
state) will be presented separately.

It is a pleasure to thank A. Morello for numerous helpful and motivating
discussions. The author is grateful to R. Sessoli and W. Wersdorfer for
providing him with complete files on the structure of the $Mn_{12}$-acetate
molecule. The author is also indebted to S. Burmistrov, L. Dubovskii and
I. Polishchuk for useful discussions. This work is supported by Russian
grant RFBR 04-02-17363a.

\begin{appendix}
\section{Two-level system}
\label{sub:apA}

The effective Hamiltonin of a biased two-level system (TLS) has the
form:
\begin{equation}
H_{TLS} = -\Delta_o \hat{\tau}^x - \xi \hat{\tau}^z,
\label{TLS.1}
\end{equation}
where $ \hat{\tau}^x, \hat{\tau}^z $ are the Pauli matrixes multiplied by 2;
$\Delta_o$ is the tunneling matrix element; and $\xi$ is the asymmetry between
two states (i.e., the longitudinal bias). One can easily solve this problem for
eigenfunctions:
\begin{eqnarray}
| \Uparrow \rangle &=& u | \uparrow \rangle + v | \downarrow \rangle; \;
| \Downarrow \rangle = -v | \uparrow \rangle + u | \downarrow \rangle; \;
\nonumber \\
(u,v) &=& ( (\varepsilon \pm \xi) / 2 \varepsilon )^{1/2}; \;\;\;
\varepsilon = \sqrt{ \xi^2 + \Delta^2_o },
\label{TLS.2}
\end{eqnarray}
where the corresponding energies in states $| \Uparrow \rangle, | \Downarrow
\rangle$ are given by $E_{\Uparrow,\Downarrow}= \mp \varepsilon$ and
\begin{equation}
| \uparrow \rangle = {{1}\choose{0}}; \;\;\;  | \downarrow \rangle =
{{0}\choose{1}}.
\label{TLS.3}
\end{equation}
If at time $t=0$ system was in state $| \uparrow \rangle$, the probabilities
to find system at time t in states $| \uparrow \rangle$ or $| \downarrow
\rangle$ are
\begin{equation}
P_{\uparrow \uparrow} = 1 - {\Delta^2_o \over \varepsilon^2} \sin^2(\varepsilon t
/ \hbar); \;\;\; P_{\downarrow \uparrow} = {\Delta^2_o \over \varepsilon^2}
\sin^2(\varepsilon t / \hbar).
\label{TLS.4}
\end{equation}
This describes the oscillations with frequency $\varepsilon$ between states
$| \uparrow \rangle$ and $| \downarrow \rangle$. In the limit $\Delta_o << \xi$
the oscillations are suppressed since their amplitude is $\Delta^2_o / \xi^2 << 1$.

\section{Method of calculations}
\label{sub:Calc_meth}

\emph{\textbf{(1) $W_D(H^{\perp})$.}} ~ To obtain the transverse field
behavior of the dipolar bias distribution half-width $W_D$ in the crystals
of $Fe_8$ and $Mn_{12}$ molecules, two clusters of different crystal
symmetry are used. {\it (a) The $Fe_8$ crystal.} The cluster for the
$Fe_8$ system contains $50^3$ unit cells arranged in a triclinic
lattice array with lattice parameters: \cite{CDC} $a=10.522(7)$ {\AA};
$b=14.05(1)$ {\AA}; $c=15.(1)$ {\AA} with angles $\alpha = 89.90(6)^o;
\beta = 109.65(5)^o; \gamma = 109.27(6)^o$. Each unit cell of volume
$V_o \approx 1969$ {\AA}$^3$ contains eight spin-$5/2$ $Fe^{+3}$ ions,
correctly positioned and oriented. \cite{CDC,OrientFe} {\it (b) The
$Mn_{12}$ crystal.} The cluster for the $Mn_{12}$ system contains
$50^3$ unit cells, arranged in a tetragonal lattice array with lattice
parameters: \cite{CDC} $a=b=17.1627(6)$ {\AA}; $c=12.2880(4)$ {\AA} with
angles $\alpha = \beta = \gamma \approx 90^o$. Each unit cell of
volume $V_o \approx 3619.5$ {\AA}$^3$ contains twelve spins: four
spin-$3/2$ $Mn^{4+}$ ions in the inner shell and eight spin-$2$ $Mn^{3+}$
ions in the outer shell, correctly positioned and oriented. \cite{CDC}

The distributions of the dipolar bias fields and energies in the cluster
are calculated taking into account all internal spins $\vec{s}^{(p)}_i$
of each molecule ($\vec{S_i} = \sum_p \vec{s}^{(p)}_i$, $p=8$ in $Fe_8$
and $p=12$ in $Mn_{12}$). The internal molecular spins $\vec{s}^{(p)}_i$
cannot flip independently - each molecule changes its total spin orientation
as a rigid object. Initially, all molecules in the sample are oriented
along the easy axis, either at random with projections $S^z_i= \pm S_i$
(for depolarized sample with initial magnetization $M=0$), or with
projections $S^z_i= + S_i$ (for polarized sample with $M=1$).

To obtain the average longitudinal bias field acting on $i$-th molecule,
we calculate the longitudinal fields $h^{z(p)}_i$, created by all internal
spins of all molecules in the sample at each $p$-th internal spin
$\vec{s}^{(p)}_i$ of $i$-th molecule. Then, the average longitudinal field
at $i$-th molecule is $H^z_i = \sum_p (s^{(p)}_i / S_i) h^{z(p)}_i$. The
calculation of the average transverse field is similar.

First, we calculate the longitudinal and transverse fields in the sample,
assuming $S^z_i = S_i$ and $S^x_i = 0$ for each molecule. Knowing internal
and external longitudinal and transverse fields at each molecule from the
first step, we calculate $\Delta_i$, $S^z_i$ and $S^x_i$ by means of exact
diagonalization of the molecular Hamiltonians (\ref{Hfe},\ref{Hmn}). Obtaining
$S^z_i$ and $S^x_i$ for each molecule, we repeat calculation of fields in the
sample. Using these new fields, we recalculate $\Delta_i$, $S^z_i$ and $S^x_i$
and so on. This iteration procedure converges and, depending on the value of
external applied field, it is enough to make $10-20$ iterations to obtain a
final result.

\begin{figure}[ht]
\centering
\vspace{-2.2cm}
\hspace{0.0cm}
\includegraphics[scale=0.4]{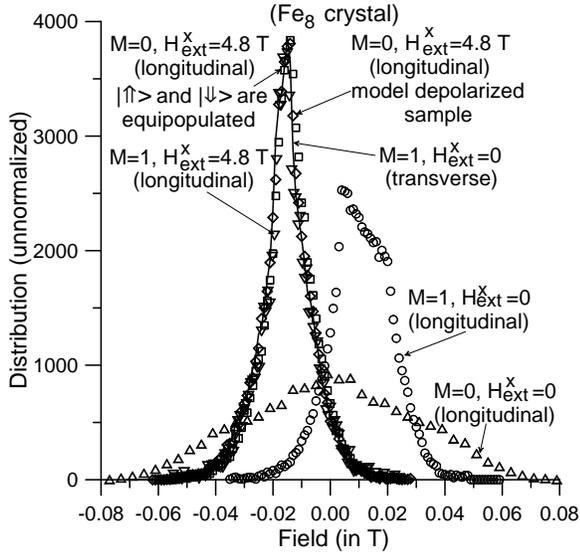}
\vspace{-1.8cm}
\caption{Total distributions of the internal dipolar {\it fields} in the
crystal of $Fe_8$ molecules (cluster of $50^3$ molecules arranged in a
triclinic lattice). The curves are: {\it solid line} - the transverse
dipolar fields distribution for $M=1$ (polarized sample) at $H^x_{ext}=0$;
{\it inverse triangles} - the longitudinal dipolar fields distribution
for $M=1$ at $H^x_{ext}=4.8 \; T$; {\it squares} - the longitudinal dipolar
fields distribution for $M=0$ at $H^x_{ext}=4.8 \; T$ (depolarized sample -
states $|\Uparrow \rangle$ and $|\Downarrow \rangle$ are equipopulated);
{\it diamonds} - the same as squares, but for the model depolarized sample
(all molecules are in states $|\Uparrow \rangle$, but $\sum_i S^z_i / |S^z_i|
= 0$); {\it circles} - the longitudinal dipolar fields distribution for $M=1$
at $H^x_{ext}=0$ (polarized sample); {\it triangles} - the longitudinal dipolar
fields distribution for $M=0$ at $H^x_{ext}=0$ (depolarized sample).}
\label{fig:DE_Fig5}
\end{figure}

\vspace{-3mm}

To show that this iteration procedure converges, we present
Fig.\ref{fig:DE_Fig5} where the total distributions of longitudinal
and transverse {\it fields} in the crystal of $Fe_8$ molecules are
plotted. In a completely polarized sample ($M=1$) at zero external
transverse field $H^x_{ext}$ the "$x$" component of the dipolar field
is $H^x_d \sim S z_{ij} x_{ij} / r^5_{ij}$ ($S^x = 0$ and $S^z=S$). At
large $H^x_{ext}$ ($S^z \to 0$ and $S^x \to S$) the "$z$" component
of the dipolar field is $H^z_d \sim S x_{ij} z_{ij} / r^5_{ij}$. Thus,
the $z$-fields distribution at large $H^x_{ext}$ and the $x$-fields
distribution in a polarized sample ($M=1$) at $H^x_{ext}=0$ should be
the same. This is what one can see in Fig.\ref{fig:DE_Fig5}. After the
iteration procedure the longitudinal fields distribution for $M=0$ at
$H^x_{ext} = 4.8 \; T$ almost coincides with the transverse fields
distribution for $M=1$ at $H^x_{ext} = 0$.

Note also that the bias field distributions in the model depolarized
sample (all molecules are in states $|\Uparrow \rangle$, but $\sum_i
S^z_i / |S^z_i| = 0$; shown by diamonds in Fig.\ref{fig:DE_Fig5})
almost coincide with the distributions in the depolarized sample
(states $|\Uparrow \rangle$ and $|\Downarrow \rangle$ are equipopulated;
shown by squares). Moreover, the bias distributions in a polarized
(shown by inverse triangles) and depolarized samples (shown
by squares and diamonds) at large fields almost coincide as well.
All this means that in order to understand the field dependence of
important parameters at large transverse fields, it is sufficient to
calculate all these parameters only in the model depolarized sample.

Repeating the iteration procedure described above for each value of
external transverse field $H^x$, we obtain: 1) the distributions of
longitudinal and transverse dipolar bias fields and energies;
2) $\langle S^z \rangle (H^x)$ and $\langle S^x \rangle (H^x)$ (the
sample average absolute values of $z$ and $x$ projections of total
spin $\vec{S}_i$); and 3) $W_D(H^x)$. All these calculations can be
done for any degree of initial polarization $M$ and for any
degrees of populations of states $|\Uparrow \rangle$ and $|\Downarrow
\rangle$; the external transverse field can be applied in any
direction in the $x-y$ plane. Finally, we would like to mention that
the results obtained for cluster of $40^3$ molecules do not change
with further cluster size increase.

\emph{\textbf{(2) $E_o(H^{\perp})$.}} All (longitudinal) hyperfine
couplings $\{ \omega^{||}_k \}$ \cite{PS96,PSSB} between the $Fe$ ($Mn$)
electronic spins $\vec{s}^{(p)}$ ($\vec{S} = \sum_p \vec{s}^{(p)}$) and
the nuclear spins $\{ I_k \}$ are assumed {\it dipolar} (with the exception
of the nuclear spin of any $Fe^{57}$ isotope in the $Fe_8$ molecule and
of the nuclear spin of any $Mn^{55}$ nucleus in the $Mn_{12}$ molecule).
The strength of the contact hyperfine interaction between the $Fe$
electronic spin and the $Fe^{57}$ nuclear spin is known \cite{FreemWat}
(the nuclear spin of the $Fe^{56}$ isotope is zero; the standard $Fe_8$
molecule contains $> 97 \%$ of the $Fe^{56}$ isotope). The strength of the
hyperfine interaction between the $Mn^{3+}$, $Mn^{4+}$ electronic spins
and the $Mn^{55}$ nuclear spin can be extracted from the recent NMR
measurements. \cite{KuboNMR}

Knowing the transverse field dependence of $\langle S^z \rangle$ and
$\langle S^x \rangle$ and the positions and moments of all nuclear spins
and $Fe^{3+}$ ($Mn^{4+}$, $Mn^{3+}$) ions in a molecule, \cite{CDC} one can
calculate all the couplings $\{ \omega^{||}_k \}$ and the half-width $E_o
(H^{\perp}) = (\sum^N_{k=1} (I_k+1) I_k (\omega^{||}_k)^2 / 3)^{1/2}$ of the
hyperfine bias energies distribution. All the necessary details of calculations
can be found in literature \cite{PS96,PSSB,STPRB,STChPh,ROSE,WWISO}).

\section{Two coupled TLS}
\label{sec:2TLS}

Consider an ensemble of the interacting two-level systems and choose
any two coupled systems described by the Hamiltonian:
\begin{equation}
H = H^{'}_1 + H^{'}_2 +
\sum_{\alpha, \beta} V^{\alpha \beta} \hat{\tau}^{\alpha}_1
\hat{\tau}^{\beta}_2; \; H^{'}_i = -\Delta_i \hat{\tau}^x_i -
\xi^{'}_i \hat{\tau}^z_i.
\label{HM1}
\end{equation}
Here both $\xi^{'}_i$ {\it do not include} the bias arising from
interactions with the second TLS (the bias $\xi^{'}_i$ is created
by all other TLS in the ensemble), and the last term describes the
interactions between two systems. It is convenient to rewrite the
Hamiltonian (\ref{HM1}) equivalently, adding the contribution coming
from the second involved TLS to each $\xi^{'}_i$. In what follows,
we limit our consideration only by $\{ \alpha, \beta \} = \{ x, z \}$.
The term describing the bias created by the second TLS at the first
one is $V^{zz} \hat{\tau}^{z}_1 \hat{\tau}^{z}_2 + V^{zx} \hat{\tau}^{z}_1
\hat{\tau}^{x}_2$ and the resulting Hamiltonian becomes:
\begin{equation}
H = H_1 + H_2 - V^{zz} \hat{\tau}^{z}_1 \hat{\tau}^{z}_2 +
V^{xx} \hat{\tau}^{x}_1 \hat{\tau}^{x}_2,
\label{HM1_1}
\end{equation}
where in both $H_1$ and $H_2$ the asymmetries $\xi_i$ now contain the
contributions from all the TLS in the ensemble.

Calculating matrix elements of the Hamiltonian $H$ in the representation
(\ref{TLS.2}), one gets (the states are ordered as
$\{ |S_1 S_2 \rangle \} = \{ |\Uparrow \Uparrow \rangle, |\Uparrow
\Downarrow \rangle, |\Downarrow \Uparrow \rangle, |\Downarrow \Downarrow
\rangle \}$):
\begin{equation}
\widetilde{H} = \widetilde{H}_0 + \widetilde{V}_{dd};
\label{HM2}
\end{equation}
\begin{equation}
\widetilde{H}_0 = \left( \begin{array}{cccc}
-\varepsilon_1 - \varepsilon_2 & 0 & 0 & 0\\
0 & -\varepsilon_1 + \varepsilon_2 & 0 & 0 \\
0 & 0 & \varepsilon_1 - \varepsilon_2  & 0 \\
0 & 0 & 0 & \varepsilon_1 + \varepsilon_2  \\
\end{array} \right),
\label{HM3}
\end{equation}
where $\varepsilon_i = \sqrt{\xi^2_i + \Delta^2_i}$ and
\begin{equation}
\widetilde{V}_{dd} = \left( \begin{array}{cccc}
  A &  C &  D & B \\
  C & -A & B & -D \\
  D & B & -A & -C \\
  B & -D & -C & A \\
\end{array} \right)
\label{HM4}
\end{equation}
with $A, B, C, D$ given by
\begin{eqnarray}
A &=& -V^{zz} {\xi_1 \xi_2 \over \varepsilon_1 \varepsilon_2} +
V^{xx} {\Delta_1 \Delta_2 \over \varepsilon_1 \varepsilon_2} \nonumber \\
B &=& -V^{zz} {\Delta_1 \Delta_2 \over \varepsilon_1 \varepsilon_2} +
V^{xx} {\xi_1 \xi_2 \over \varepsilon_1 \varepsilon_2} \nonumber \\
C &=& V^{zz} {\xi_1 \Delta_2 \over \varepsilon_1 \varepsilon_2} +
V^{xx} {\Delta_1 \xi_2 \over \varepsilon_1 \varepsilon_2} \nonumber \\
D &=& V^{zz} {\Delta_1 \xi_2 \over \varepsilon_1 \varepsilon_2} +
V^{xx} {\xi_1 \Delta_2 \over \varepsilon_1 \varepsilon_2}.
\label{HM5}
\end{eqnarray}
In the limit $\Delta_i >> \xi_i$ matrix elements $C$ and $D$ are small
in comparison with $A$ and $B$.

Note that states $|\Uparrow\rangle$ and $|\Downarrow\rangle$ are the
eigenstates of the Hamiltonian (\ref{TLS.1}) and in both TLS only
transitions between these states (but not between states
$|\uparrow\rangle$ and $|\downarrow\rangle$ (\ref{TLS.3})) are considered.
Two central states $| \Uparrow \Downarrow \rangle$ and $| \Downarrow \Uparrow
\rangle$ (from now on we call them the "flip-flop states") are separated from
the two remaining states $|\Uparrow \Uparrow \rangle$ and $|\Downarrow
\Downarrow \rangle$ by the energy gaps $ > \Delta_i$ and in the limit $\Delta_i
>> \{ |\xi_i|, |V^{\alpha \beta}| \}$ the effect of these two remaining
states on the flip-flop transitions $|\Uparrow \Downarrow \rangle
\Leftrightarrow |\Downarrow \Uparrow \rangle$ is small (within a
second-order perturbation theory the corrections to the flip-flop matrix
elements are $ \sim (V^{\alpha \beta})^2 \xi^2_i/\Delta^3_i$). Therefore,
in this limit two flip-flop states of a pair can be considered as an
effective two-level system. The coefficient $A$ in (\ref{HM5}) plays
a role of the mean energy for this effective TLS.

In the limit $\Delta_i >> \{ |\xi_i|, |V^{\alpha \beta}| \}$ the tunneling
matrix element describing the flip-flop transitions is given by $|B|$; the
energy change during these transitions is $2|\varepsilon_1-\varepsilon_2|
<< \Delta_i$. Then the probability to find system at time $t$ in state
$|\Downarrow \Uparrow \rangle$ if at $t=0$ it was in state $|\Uparrow
\Downarrow \rangle$ is:
\begin{equation}
P_{(\Downarrow \Uparrow) (\Uparrow \Downarrow)} \sim
{\Delta^2_{ff} \over E^2_{ff}} \sin^2(E_{ff} t / \hbar),
\label{FF.1}
\end{equation}
where the frequency of oscillations, the tunneling matrix element connecting
states $|\Uparrow \Downarrow \rangle$ and $|\Downarrow \Uparrow \rangle$ and
the asymmetry between these two states are given by:
\begin{equation}
E_{ff} \sim \sqrt{\xi^2_{ff} + \Delta^2_{ff}};
\label{FF.2}
\end{equation}
\begin{equation}
\Delta_{ff} \sim |B|; \;\;\;
\xi_{ff} \approx |\varepsilon_1 - \varepsilon_2|.
\label{FF.3}
\end{equation}
If $\xi_{ff} < \Delta_{ff}$, the amplitude of oscillations between states
$|\Uparrow \Downarrow \rangle$ and $|\Downarrow \Uparrow \rangle$ is
$\Delta^2_{ff} / E^2_{ff} \sim 1$. Since all other transitions are described
by the matrix elements $\sim |V^{\alpha \beta}| \xi_i/\Delta_i$ and the
energy change during these transitions is $> \Delta_i$, in the limit
$\Delta_i >> \{ |\xi_i|, |V^{\alpha \beta}| \}$ all these transitions
are not in resonance and their amplitude is small.

One note is in order here. Applying this model at low temperatures to the
central molecular spins $\vec{S}_i$ described by the Hamiltonians of magnetic
anisotropy (\ref{Hfe},\ref{Hmn}), one should take into account that the
spin projection $S^{\perp}$ along the transverse magnetic field $H^{\perp}$
has the same sign in both exacts lowest states $|\Uparrow \rangle$ and
$|\Downarrow \rangle$ of (\ref{Hfe}) and (\ref{Hmn}). The eigenstates of
these Hamiltonians can be easily obtained by the exact diagonalization
method.

\section{Flipon "random walks"}
\label{sub:Rand_W}

Consider quasi-$1d$ (i.e., along one crystallographic axis) motion of
one flipon (the flipon center of mass moves in a "fictitious" lattice
whose sites are placed directly between the nearest-neighbor sites of
a real lattice) and let us approximate this motion by a "random walks"
model. In this approximation we assume that flipon can: (i) make a jump
to the right with the probability $p$; (ii) make a jump to the left
with the probability $q$; and (iii) stay at the same site with the
probability $s=1-p-q$. The probability to find a "walk" with the $K_r$
lattice steps to the right and $K_l$ steps to the left from total N
steps is given by the {\it polynom} distribution
\begin{equation}
P_N(K_r,K_l) = {N! \; p^{K_r} q^{K_l} s^{N-K_r-K_l} \over
K_r! K_l! (N-K_r - K_l)!};
\label{Polyn1}
\end{equation}
\begin{equation}
\sum^N_{K_r=0} \sum^{N-K_r}_{K_l=0} P_N(K_r,K_l) = (p+q+s)^N=1.
\label{Polyn2}
\end{equation}
The corresponding displacement for such walk is $\widetilde{r} =
K_r-K_l$. The average displacement $\langle \widetilde{r} \rangle$
and dispersion $\langle (\Delta \widetilde{r})^2 \rangle =
\langle \widetilde{r}^2 \rangle - \langle \widetilde{r} \rangle^2$
can be easily calculated
\begin{equation}
\langle \widetilde{r} \rangle = N (p-q); \;\;\;
\langle (\Delta \widetilde{r})^2 \rangle = 4 N p q + N s (1-s)
\label{displac}
\end{equation}
and for $p=q=s=1/3$ one gets $\langle \widetilde{r} \rangle = 0$ and
$\langle (\Delta \widetilde{r})^2 \rangle = 2 N / 3$. In our case
the time $t = N \Delta \tau$ is measured in units of $\Delta \tau =
\hbar / \Delta^{nn}_{ff}(\widetilde{a})$ and the distance (the
displacement) $r = \widetilde{r} \Delta r$ is in units of $\Delta r =
\widetilde{a}$ (these parameters describe the flipon elementary "jump").
The distribution $P_N(K_r,K_l)$ defines the normalized distribution
$P_N(\widetilde{r})$ ($\widetilde{r} =  K_r-K_l \in [-N,N]$). For large
values of $N$, this distribution transforms into the Gaussian
one $P_1(r,t) = \exp (-r^2 / 4 D_f t) / (4 \pi D_f t)^{1/2}$, where
$D_f$ is the flipon effective diffusion coefficient
\begin{equation}
D_f = \rho \widetilde{a}^2 \Delta^{nn}_{ff} / 2 \hbar; \;\;\; \rho
= 4 p q + s (1-s)
\label{FDC}
\end{equation}
and $\langle r^2(t) \rangle = 2 D_f t$. The Gaussian $P_1(r,t)$ gives
the probability to find a flipon at time $t$ at the distance $r$ from its
$t=0$ position.

\end{appendix}


\begin{thebibliography}{99}
\bibitem{DeltaDistr} Some molecular systems contain a minor molecular species
(randomly distributed in a sample of major species) having different
values of $\Delta_o$. Often these species are not characterised at
all and in the present work we will not concentrate on this effect.
Thus, in this work we mostly consider a sample of identical
molecules (although impurities can change the width of the dipolar
fields distribution).
\bibitem{FeExpl} These parameters were chosen \cite{TB} to reproduce both the
value of zero-field tunneling splitting \cite{WWPRL} and the period of oscillations
of $\Delta_o(H^{\perp})$ in a transverse field. \cite{WWSSCI}
\bibitem{SxSign} Note that at large transverse fields the spin projection
$S^x_i$ in both exact lowest states $|\Uparrow_i \rangle$ and $|\Downarrow_i
\rangle$ of the Hamiltonians (\ref{Hfe}) and (\ref{Hmn}) has the same sign
and $S^z_i \to 0$.
\bibitem{RES_EX} It is worth mentioning that the calculated values of these
parameters and those measured in both systems experimentally
\cite{WWPRL,WWISO,WWEPL} are in close agreement.
\bibitem{NatIsot} Some natural isotopes (like, for example, $Fe^{56}$ with the
natural concentration $> 97 \%$ and $H^1$ with the natural concentration
$\approx 99.985$ in the $Fe_8$ molecule) can be chemically replaced by other
isotopes (like $Fe^{57}$ and $H^2$).\cite{WWISO}
\bibitem{2FDF} The pair distribution function is approximated by a
product $P_N(\widetilde{r}_1) P_N(\widetilde{r}_1)$ since in the limit
$\Delta^{nn}_{ff} >> \xi^{nn}_{ff}$ at $R_{ex} > 2 \widetilde{a}$
each excited spin has equal probabilities to create a resonance with
both its neighbors along the same axis and two flipons move almost
independently.
\bibitem{3drw} In a continuous $3d$ limit we get $\langle r^2(t) \rangle
= 6 D^{(3)}_f t$, where $D^{(3)}_f = \rho \widetilde{a}^2 / 6 t_f$ and
$\rho$ is also defined by Eq.(\ref{FDC}).
\end{thebibliography}
\end{document}